# Vacancy-Induced Boron Nitride Monolayers as Multifunctional Materials for Metal Ion Batteries and Hydrogen Storage Applications

Wadha Al Falasi[1,2], Wael Othman[3,4], Tanveer Hussain[5], and Nacir Tit[1,2*]

[1]Physics Department, UAE University, Al Ain, United Arab Emirates

[2]Water and Energy Research Center, UAE University, Al-Ain, United Arab Emirates

[3]Engineering Division, New York University Abu Dhabi, Abu Dhabi, United Arab Emirates

[4]Mechanical and Aerospace Engineering, Tandon School of Engineering, New York University, NY 10012, USA

[5]School of Science and Technology, University of New England, Armidale, New South Wales, 2351, Australia

## Abstract

Energy storage through metal-ion batteries (MIBs) and hydrogen ($H_2$) fuel presents significant opportunities for advancing clean energy technologies. This study comprehensively examined the structural, electronic, electrochemical, and energy storage properties of boron-vacancy induced porous boron nitride monolayers (BN:$V_B$) as multifunctional materials, anodes for MIBs and $H_2$ storage applications. Our computational approaches, density functional theory (DFT), ab initio molecular dynamics (AIMD), and thermodynamic analysis, revealed exceptionally high energy and gravimetric densities for MIBs and $H_2$ storage, respectively. We investigated the interactions of Li, Na, and K atoms on BN:$V_B$, which strongly bonded with binding energies stronger than their bulk cohesive energies, which ensured structural stability and the absence of metal clustering. Electronic properties, analyzed through spin-polarized partial density of states (PDOS), band structure, and Bader charge analysis, revealed significant charge transfers from the metal atoms to BN:$V_B$, enhancing the electronic conductivity of the latter. Theoretical specific capacities were calculated as 1821.53, 786.11, and 490.51 mA h/g for Li, Na, and K, respectively, which comfortably exceeded the conventional anodes, such as graphite. Average open-circuit voltages (OCVs) were found as 0.15, 0.25, and 0.32 V, for Li, Na, and K, respectively, indicating strong electrochemical stability. Diffusion studies showed lower barriers of 0.47, 0.08, and 0.60 eV for Li, Na, and K, respectively, with increased metal loadings, suggesting enhanced mobilities and charge/discharge rates. On the other side, the metal-functionalized BN:$V_B$ monolayers exhibited remarkably high $H_2$ gravimetric capacities of 10.64, 10.72, and 9.38 wt% for 4Li-,4Na-, and 4K@BN:$V_B$, respectively, all surpassing the 5.50 wt% target set by the US Department of Energy for 2025. Average adsorption energies of $H_2$ on 4Li-, 4Na-, and 4K@BN:$V_B$, were found in perfect range for practical storage applications. The potentials for practical $H_2$ storage were further supported by Langmuir adsorption model-based statistical thermodynamic analysis, which examined the adsorption and desorption behavior of $H_2$ under practical conditions. These findings position BN:$V_B$ as a promising multifunctional candidate for high-performance MIBs anodes and $H_2$ storage material.

(*) Corresponding author: ntit@uaeu.ac.ae

## 1. Introduction

The development of novel materials for energy storage plays a pivotal role in meeting the worldwide need for sustainable, ecofriendly, and effective energy solutions [1]. Metal-ion batteries (MIBs), particularly lithium-ion batteries (LiIBs), are fundamental to the operation of modern electronic devices and electric vehicles due to their high energy density, straightforward functionality, and long lifespan. The performance of MIBs batteries is highly dependent on the properties of the anode materials, which play a crucial role in determining the efficiency and longevity. Traditional anode materials face limitations in capacity and rate performance. For example, graphite, the conventional anode material for LiIBs, has a limited theoretical capacity of 372 $mAhg^{-1}$, which is insufficient to meet the demands of next-generation batteries [2], [3]. As a result, there is an ongoing search for the novel materials that can offer higher capacity, better rate performance, and enhanced stability.

Alternative materials, such as silicon, tin, and various forms of carbon, which exhibit higher theoretical capacities, have been explored. However, most of them often suffer from significant volume expansion during cycling, leading to rapid capacity fading and mechanical failure [4], [5], [6], [7]. In addition, the growing need for efficient energy storage systems, fueled by the popularity of portable electronics and electric vehicles, has underscored the limitations of LiIBs [8]. Concerns such as scarcity, high cost, and safety issues associated with lithium (Li) have prompted investigations into alternative battery systems such as sodium-ion (NIBs), potassium-ion (KIBs), magnesium-ion (MgIBs), calcium-ion (CaIBs), and zinc-ion batteries (ZnIBs) [9], [10], [11], [12], [13], [14], [15]. These systems offer distinct advantages, such as abundance and wide distribution of raw materials, reduced costs, and sustainability. For example sodium (Na) and potassium (K) ranks as the sixth and seventh most abundant element in earth's crust, respectively, making NIBs and KIBs promising alternatives [9], [13]. Moreover, the mentioned systems are environmentally friendly and safer; MgIBs and ZnIBs, in particular, offer improved safety profiles due to their lower operating voltages and reduced reactivity with electrolytes [14], [16].

However, transforming to alternative MIBs presents specific challenges. A major concern is attributed to achieving electrochemical performance comparable to LiIBs, as NIBs typically exhibit lower energy densities owing to the larger size of $Na^+$ ions, resulting in reduced voltage and capacity [17], [18], [19]. Additionally, finding suitable electrode materials capable of effectively intercalating these metal atoms is intricate; for instance, graphite is inadequate for NIBs, necessitating exploration of alternatives like hard carbon and titanium-based compounds [20], [21]. Electrolyte stability and compatibility are also critical; MgIBs encounter issues with passivation layers that hinder ion movement [22], [23]. Furthermore, ensuring durability and efficiency over long cycles remains a challenge, evident in KIBs that undergo significant volume changes during use [24]. Overcoming these technical hurdles through ongoing research and development is essential to unlock the full potential of alternative MIBs [25].

Hydrogen ($H_2$), on the other hand, is the simplest molecule possessing much higher energy density per unit mass than any other fuel and yield only water byproduct upon its usage in the fuel cell [26]. Despite these advantages, the efficient storage of $H_2$ remains a technological hurdle towards adoption of $H_2$-based economy. Traditional storage approaches, such as high-pressure gas cylinders and cryogenic liquid $H_2$ tanks, pose safety and efficiency concerns. Consequently, solid-state $H_2$ storage materials, including metal hydrides, complex hydrides, and porous materials, are under extensive investigation to achieve high $H_2$ storage capacity at practical pressure and temperature conditions [38]. To address the storage challenge, the US Department of Energy (DOE) has set specific targets for onboard $H_2$ storage systems, aiming for 5.5 wt % and 40 g/L by 2025 [27]. Over the last two decades, porous solid-state materials and their composites have emerged as promising contenders for $H_2$ storage due to their unique attributes, such as customizable pore structures, large surface areas, and uniform porosity. Among these materials, metal-organic frameworks (MOFs) [28], [29], covalent organic frameworks (COFs) [30], [31], porous organic polymers (POPs) [32], and carbon-based materials [33], [34] have attracted extensive attention for $H_2$ storage applications. Despite these efforts, the limited storage capacity underscores the urgent need to explore alternative materials that can efficiently adsorb and release $H_2$ under ambient conditions [34], [36], [37]. Two-dimensional (2D) materials have garnered significant research interest because of their distinctive characteristics, such as large surface areas, adjustable porosity, and the capacity to incorporate diverse chemical functionalities. These features make them exceptionally beneficial for numerous applications, especially in catalysis, adsorption, separation processes, and energy storage [39].

Boron nitride (BN), a compound belonging to the III-V group, is renowned for its diverse structural forms and intriguing properties. One of its most prominent variations is the $sp^2$-hybridized layered BN, characterized by honeycomb-shaped monolayers composed of boron (B) and nitrogen (N) atoms, similar to the arrangement found in graphene [40]. The intrinsic properties of pristine BN are largely defined by its atomic structure, where B and N atoms are alternately arranged in a hexagonal lattice. This configuration results in a wide band gap, typically ranging around 5.0 – 7.0 eV [40], [41], [42], [43], [44], making BN an excellent electrical insulator. This wide band gap is beneficial for applications requiring high dielectric strength and thermal stability. Moreover, the strong $sp^2$ hybridization of the B-N bonds endows the material with excellent mechanical properties, including high tensile strength and flexibility [41]. However, the functional versatility of BN monolayers can be greatly enhanced by introducing atomic-scale modifications, such as vacancies, doping, and the adsorption of foreign atoms or molecules. One of the most extensively studied modifications is the creation of B vacancies ($BN:V_B$). The removal of B atoms from the lattice creates localized states within the band gap that can act as traps for charge carriers or as reactive sites for chemical functionalization. Such modifications can transform the insulating BN monolayer into a material with tunable electronic properties, thereby broadening its potential applications [45].

In recent developments, the synthesis of a novel two-dimensional porous carbon structure, known as holey graphyne (HGY), has garnered significant attention. This structure, characterized by a periodic arrangement of six and eight-vertex rings, exhibits promising semiconductor properties with high electron and hole mobility at room temperature [46]. Furthermore, the uniform periodic holes within the HGY framework offer potential for energy storage applications. Researchers have investigated $H_2$ storage capabilities within HGY structures decorated with Li and scandium (Sc), indicating promising opportunities for $H_2$ storage [47] [48]. Building upon these advancements, Mahamiya et al. [48] explored the possibility of synthesizing a BN analog of the two-dimensional HGY structure. Through theoretical investigations, they have established the energetic and dynamic stability of the porous BN (Pr-BN) structure, suggesting its suitability for various applications. They reported exceptional catalytic activity of the Pr-BN structure towards the $H_2$ evolution reaction (HER).

Given these attributes, the proposed Pr-BN system holds promise for a range of applications in nano electronics, optoelectronics, solar cells, and particularly for $H_2$ production and storage. The unique combination of stability, tunability, and catalytic activity makes Pr-BN an attractive candidate for MIBs electrode and $H_2$ storage. Its large surface area provides abundant sites for adsorption, while its chemical stability ensures durability under operational conditions. Moreover, the porous nature of Pr-BN can be tailored to optimize both uptake capacity and release or diffusion processes, rendering it a versatile material for this application [39], [48], [49].

## 2. Computational Methods

In this study, we utilized spin-polarized calculations to study the structural, electronic, and metal adsorption properties, as implemented in the "Vienna Ab-initio Simulation Package" (VASP) within the frameworks of DFT and the projected augmented wave (PAW) method [50]. The exchange and correlation interactions were addressed using Perdew-Burke-Ernzerhof form of the generalized gradient approximation (PBE-GGA) functional [51]. The van der Waals (VdW) interactions were tackled using DFT-D3 method of Grimme [52]. For structural optimization, the convergence tolerances for total energy of $10^{-8}$ eV and atomic force of 0.01 eV/ Å were applied. In sampling the Brillouin zone, we utilized a Monkhorst-Pack technique [53] with k-mesh grid of size 5×5×1 for total energy calculation. A denser k-mesh grid of size 8x8x1 was used for the density of states calculations (DOS) and a plane-wave cut-off energy of 500 eV was utilized. Bader-charge analysis within the framework of VASP package was explored to calculate the charge transfer [54].

The average binding energy ($E_{bind}$) of metal atoms (MA) embedded in the pores of Pr-BN/BN:$V_B$ monolayer was calculated using the formula:

$$E_{bind} = \frac{E_{BN+nMA} - (E_{BN} + nE_{MA})}{n} \quad (1),$$

where $E_{BN+nMA}$, $E_{BN}$, and $E_{MA}$ stand for the total energies of the system of Pr-BN/BN:$V_B$ with and without the $n$ MA, and isolated MA, respectively. Furthermore, ab initio molecular dynamics (AIMD) simulations were conducted to confirm the structural stability of MA-functionalized BN:$V_B$ at 400 K. Temperature control was achieved using the Nose-Hoover thermostat, employing a time step of 1.0 fs over an 8 ps simulation period.

The theoretical specific capacity of the anode material is calculated by [55]:

$$C_Q = \frac{xF}{3.6M_{(nMA@BN:V_B)}} \quad (2),$$

where $x$ is the maximum number of electrons involved in the electrochemical process, $n$ is the maximum number of adsorbed MA ($x = n$ for Li, Na, and K atom intercalation/de-intercalation processes), $F$ is the Faraday constant with a value of 96485.3 C mol$^{-1}$, and $M_{(nMA@BN:V_B)}$ is the mass of the fully functionalized $BN: V_B$ in g/mol ($M_{(nMA@BN:V_B)} = n_{MA}M_{MA} + n_B M_B + n_N M_N$).

The open-circuit voltage (OCV) is expressed by [56]:

$$OCV = \frac{E_{BN:V_B} - E_{nMA@BN:V_B} + nE_{MA}}{ne} \quad (3),$$

where $E_{BN:VB}$, $E_{nMA@BN:VB}$, and $E_{MA}$ stand for the total energies of the BN:$V_B$ before and after functionalization with MA, and isolated MA, respectively. $n$ is the total number of adsorbed MA (Li, Na, or K), and the denominator $ne$ is the valency of the MA.

The average adsorption energy ($E_{ads}$) of $H_2$ molecules on top of the substrate is defined as:

$$E_{ads} = \frac{E_{4MA@BN:V_B+mH_2} - (E_{4MA@BN:V_B} + mE_{H_2})}{m} \quad (4),$$

where $E_{4MA@BN:V_B+mH_2}$, $E_{4MA@BN:V_B}$, and $E_{H_2}$ represent the total energies of 4MA@BN:$V_B$ with and without the $m$ $H_2$ molecule(s), and isolated $H_2$ molecule, respectively.

The gravimetric $H_2$ uptake capacity is defined as follows [32]:

$$C_T(wt\%) = \left[\frac{N_T \cdot M(\mathrm{H_2})}{N_T \cdot M(\mathrm{H_2}) + N_{MA} \cdot M(\mathrm{MA}) + N_B \cdot M(\mathrm{B}) + N_N \cdot M(\mathrm{N})}\right] \times 100\% \quad (5),$$

where $M(\mathrm{H_2})$, $M(\mathrm{MA})$, $M(\mathrm{B})$, and $M(\mathrm{N})$ stand for the molecular masses of $H_2$ molecule, MA, and host crystal atoms (i.e., B and N), respectively, whereas $N_T$, $N_{MA}$, $N_B$, and $N_N$ stand for the total number of $H_2$ molecules, MA, B, and N in the system, respectively.

The adsorption-desorption behaviors under operational conditions were analyzed using statistical thermodynamics based on the grand canonical partition function ($z$) [57]:

$$z = 1 + \sum_{i=1}^{n} e^{-\frac{(E_b^i - \mu)}{k_B T}} \quad (6),$$

where $n$ represents the maximum number of $H_2$ molecules adsorbed on the substrate. $E_b^i$, $k_B$, $T$, and $\mu$ represent $E_{ads}$ of the $i^{th}$ $H_2$ molecule, the Boltzmann constant (1.38 x $10^{-23}$ J/K), the absolute temperature, and the chemical potential of the $H_2$ molecule in the gas phase, respectively. In particular, $\mu$ is a function of $P$ and $T$, is defined as:

$$\mu_{H_2}(P,T) = \Delta H + T\Delta S + k_B T ln \frac{P}{P_0} \quad (7),$$

where $\Delta H$, $\Delta S$, $P$, and $P_0$ represent the enthalpy and entropy change, pressure, and the atmospheric pressure of 1.01 x $10^5$ Pa, respectively. The values of $\Delta H$ + $T\Delta S$ are obtained from the experimental database [58]. Meanwhile, the number of stored $H_2$ molecules ($N$) in the host material can be expressed by:

$$N = N_0 \left[\frac{Z-1}{Z}\right] \quad (8),$$

where $N_0$ represents the maximum number of $H_2$ molecules adsorbed on the host medium at 0 K ($N_T$).

# 3. Results and Discussions

## 3.1 Structural and Electronic Properties

The system used in our study consists of the primitive cell of Pr-BN. The optimized structure shows a periodic pattern of six and eight-vertex rings, similar to HGY [49]. The Pr-BN has a structure of Bravais square lattice which belongs to the space group $C_5^1$ (Pm #6) and its primitive cell (PC) contains 12 B and 12 N atoms as shown in Figure 1(i-a). The bond length between B and N atoms ranges between 1.30 to 1.49 Å. The range of different values of BN bond lengths attribute to a combination of three different types of bonding reported in the case of Pr-BN, $sp^2$-$sp^2$, $sp^2$-sp, and sp-sp hybridized B and N atoms [49]. The lattice constants of Pr-BN are a= b= 10.96 Å, consistent with the values reported by Mahamiya et al. (a= b= 10.92 Å) [49]. Figure 1(i-b) shows that Pr-BN exhibit a nonmagnetic semiconductor characteristics with wide band gap of 4.21 eV, which is consistent with the literature [49]. Dąbrowska et al. reported that B vacancy ($V_B$) is always formed during hexagonal BN (h-BN) synthesis, but with different concentrations and have shown more stability compared to N vacancy ($V_N$) [45]. The existence of $V_B$ enhances the catalytic activity of h-BN [59]. Hence, to improve the catalytic activity and to tailor the properties of Pr-BN, we considered the effects of $V_B$ (i.e., BN:$V_B$). The relaxed structure of BN:$V_B$ is shown in Figure 1(ii-a). The top view of the structure demonstrates the inward relaxation and the shift from an eight-vertex ring to a seven-vertex ring with N-N bond length of 1.35 Å. The side view of the relaxed structure reveals that the B-N-B bond within the seven-membered ring is elevated by 1.61 Å. This observation aligns with previous studies on defect-induced structural modifications, where vacancies and defects are known to cause significant local distortions and reconstructions in the lattice [60], [61]. We observe a ferromagnetic characteristic in BN:$V_B$ with a magnetization of 1 $\mu_B$. The band structure in Figure 1(ii-b) reveals a new dispersive band above the Fermi level, and the band gap is reduced to 3.14 eV for spin-up electrons and 1.18 eV for spin-down electrons.

## 3.2 Metal decoration of BN:$V_B$ monolayer (MA@BN:$V_B$)

Next, the adsorption mechanism of MA over BN:$V_B$ is discussed. The specific pore structure of BN:$V_B$ can facilitate the accommodation and distribution of multiple MA. Additionally, the surface chemistry of BN:$V_B$ and the charge transfer from the MA can be engineered to enhance the binding affinity. These combined factors make BN:$V_B$ highly suitable for the effective and stable embedment/decoration of MA, enabling their use in MIBs and $H_2$ storage.

### 3.2.1 Structure and stability of MA@BN:$V_B$

For validation, Pr-BN is initially investigated as a benchmark. The adsorption of MA on Pr-BN is studied by examining the most stable and energetically favorable configurations among the five different initial structures, see Figure 2(a). Following the geometrical optimizations, it is observed that a single MA (Li, Na, K) tends to bind on top of site X1, exhibiting the lowest $E_{bind}$ according to equation (1). However, it's noteworthy that the resulting minimum $E_{bind}$ of MA on Pr-BN is weaker than the cohesive energy ($|E_{bin}| < |E_{coh}|$), as seen in Figure 2(b). Consequently, we further investigated the adsorption of MA on BN:$V_B$, as the $V_B$ was experimentally characterized and theoretically predicted to serve as an active adsorption site [59].

Like the case of Pr-BN, we considered all the available adsorption sites on BN:$V_B$. The single MA prefers to bind inside the pore (site X1), forming bonds with the surrounding N atoms of BN:$V_B$ with average bond lengths of 2.0, 2.37, and 2.86 Å for Li-, Na-, and K@BN:$V_B$, respectively. The calculated $E_{bin}$ values of -4.45, -3.79, and -4.27 eV for Li, Na, K, respectively, are much stronger than their bulk $E_{coh}$ indicating strong MA bindings and the negligible possibility of metal clustering. Subsequently, we sequentially introduced further MA, in step-wise manner, on BN:$V_B$. The first adsorbed layer comprises four MA with average $E_{bin}$ values per dopants of -2.95, -2.17, and -2.23 eV for 4Li-, 4Na-, and 4K@BN:$V_B$, respectively, as illustrated in Figure 3(i). This 4MA-layer serves as our accumulating step to reach the maximum specific capacity of BN:$V_B$ (i.e., 44Li, 28Na, and 20K), as shown in Figure 3(ii). Further details on these results are provided in the subsequent subsections.

Based on Figure 3(i), the three 4MA@BN:$V_B$ systems retain their structural integrity without any noticeable deformations, implying excellent structural stabilities. Figure 4(a) shows the average-absolute values of $E_{bind}$ of a sequence of *n*MA decorated on BN:$V_B$ (with *n* = 1–4) and compared to the $E_{coh}$ of the bulk MA. The fact that $|E_{bind}| > |E_{coh}|$ for all the three systems reveals that the MA are more inclined to bind to the lattice rather than clustering. This fosters the stability of MA@BN:$V_B$, thereby increasing their viability for MIBs anodes and $H_2$ storage. It is worth mentioning that the average $E_{bind}$ simulates the experimental synthesis of functionalization more realistically than the recursive energy, $E_{rec}$, as shown in our previous work [57]. Next, the thermodynamic stability of 4MA@BN:$V_B$ is conducted using AIMD kinetics at 400 K over a period of 8 ps with step of 1 fs. Despite the buckling in the atomic structures of BN:$V_B$ after MA functionalization, the AIMD analysis indicates insignificant fluctuations in the total energies throughout the simulation, as shown in Figure 4(b). This further validates the structural integrity of 4MA@BN:$V_B$.

### 3.2.2 Electronic properties

The cycling stability and specific capacity of MIBs electrode materials are highly correlated with electrical conductivity. The strong binding interactions between adsorbent and adsorbate can be manifested in the electronic properties. Accordingly, the electronic properties of BN:$V_B$ and MA@BN:$V_B$ are studied in terms of spin-polarized partial density of states (PDOS) and band structure plots. The PDOS plots for *n*Li@BN:$V_B$, *m*Na@BN:$V_B$, and *l*K@BN:$V_B$ are illustrated in Figure 5, where *n, m*, and *l* represent the total number of Li, Na, and K atoms, respectively. Based on Figure 1(b), the main effects of $V_B$ on the electron density of states appear near the Fermi region. Additionally, the existing asymmetry between the spin-up and the spin-down PDOS suggest magnetic properties and will be elaborated as follows. **(a)** For the *n*Li@BN system, the Li(s) orbitals interact with the B(p) and N(p) orbitals in both the valence and conduction bands. A significant shift in PDOS is observed after 4Li decoration compared to BN:$V_B$. Moreover, 4Li@BN:$V_B$ preserves its magnetic semiconductor character with a reduced bandgap of 0.17 eV and 1.42 eV for spin-up and spin-down states, respectively. The system exhibits a drastic change upon reaching the maximum specific capacity of 44Li atoms and transforms to nonmagnetic metal, suggesting that MA adsorption influences the electronic and magnetic properties of BN:$V_B$ by passivating all dangling bonds and consequently the magnetization disappears. Similar trends in the electronic properties are observed in the cases of **(b)** *m*Na@BN:$V_B$ and **(c)** *l*K@BN:$V_B$ systems as shown in Figure 5. Thus, the spin-polarized PDOS demonstrate the impact of MA decoration on the electronic properties of BN:$V_B$. The effect of these interactions is further studied through Bader charge analysis, which reveals that electrical charges transfer from the MA to BN:$V_B$. For instance, in the case of 4Li@BN:$V_B$, the Li atoms transfer 0.52 e/Li to BN:$V_B$. Similarly, in 4Na@BN:$V_B$ and 4K@BN:$V_B$, 0.54 e/Na and 0.60 e/K are transferred to BN:$V_B$, respectively. This quantitative charge transfer analysis implies that Li, Na, and K dopants act as charge donors, while BN:$V_B$ serves as a charge acceptor. The significant charge transfer results in strong binding interactions between the MA and the BN:$V_B$. Additionally, the transition of BN:$V_B$ from a semiconductor to metal, following MA functionalization, is driven by the substantial charge transfer. Table 1 summarizes the electronic properties (i.e., Bader charge transfer, band-gap energy, Fermi energy, and total magnetization) of Pr-BN, BN:$V_B$, and *n*MA@BN:$V_B$.

The metallic character is crucial for the performance of anode materials in MIBs. A metallic anode provides high electronic conductivity, which is essential for efficient charge transport during battery operation. Moreover, the substantial charge transfers transform the MA into cations, which facilitate appropriate $E_{ads}$ of $H_2$ molecules. Meanwhile, the spin-polarization calculations provide information about the magnetization, which contributes to the interaction with $H_2$ molecules in all three cases, with total magnetization of 1.00 $\mu_B$. Further details are provided in section 3.3.

We further computed the electron localization function (ELF) to assess the nature of interactions in the real space with a value between 0 and 1 (ranging from ionic to covalent bonds, respectively). First, all the three cases have shown a tendency towards forming an ionic bond with K showing the strongest ionic bond character. A strong covalent feature is found for B–N and N–N bonds, as indicated by the ELF value of 0.94 in Figure 6. However, the ELF values decrease to zero at the center of the pores, especially the larger pore, implying a low electron distribution. This is crucial for diffusion channel formation and MA adsorption.

### 3.3 MIBs anodes using nMA@BN:$V_B$

#### 3.3.1 Theoretical specific capacity and open-circuit voltage

The assessment of specific capacity and open-circuit voltage (OCV) holds significant importance in evaluating the electrochemical behavior of electrodes in MIBs [62]. To model the electrochemical process and ascertain the maximum specific capacity, we have systematically introduced Li, Na, and K atoms to BN:$V_B$, incrementally adding four MA (i.e., 1 layer) at a time. The average $E_{bind}$ surpassing the $E_{coh}$ is considered as an indication for the further adsorption of MA. It is evident that MA exhibited unique adsorption characteristics attributable to their diverse atomic sizes and $E_{bind}$. Owing to its high surface-to-volume ratio, the BN:$V_B$ could host up to 44 Li, 28 Na, and 20 K. The average $E_{bind}$ are 1.80, 1.36, and 1.26 eV per Li, Na, and K, respectively, as shown in Figure 7(a). It is crucial to note that the structure will exhibit symmetry breaking upon reaching the maximum specific capacity, hence allowing orbitals mixing as shown in Figure 3(ii). Next, the theoretical specific capacities are determined using equation (2). The Li, Na, and K specific capacities of BN:$V_B$ anode are calculated as 1821.53, 786.11, and 490.51 mA $hg^{-1}$, comfortably exceeding that of BSi (1374 $mAhg^{-1}$ for Li) [63], graphite (372 $mAhg^{-1}$ for Li) [64], 2D $PC_3$ (1200 $mAhg^{-1}$ for Li) [65] , $Zn_{0.8}Mn_{0.2}O$ (312.4 $mAhg^{-1}$ for Na) [66], $V_2N$ MXenes (952 $mAhg^{-1}$ and 463 mA h/g for Na) [67], T-$NiSe_2$ (247.0 $mAhg^{-1}$ for K) [68], and $ZnSiP_2$ (517.0$mAhg^{-1}$ for K) [69]. In addition, the OCV is calculated using equation (3). As shown in Figure 7(b), the average OCVs of BN:$V_B$ anodes are 0.15, 0.25, and 0.32 V for Li, Na, and K, respectively. These optimal OCV values are desirable for efficient anode materials to prevent dendrite formation during cycling and safeguard MIBs from short lifespan and breakdown.

#### 3.3.2 Diffusion barriers

To further assess the dynamics of MA on the surface of BN:$V_B$, we have computed the diffusion energy barriers for the movement of MA from the most stable position to a neighboring equivalent position. Strong $E_{bind}$ values in the cases of single Li, Na, and K adsorption suggest significantly high diffusion barriers. This assumption is further corroborated by the diffusion barrier calculations presented in Figure 8. The diffusion barriers for Li-, Na-, and K@BN:$V_B$ are found as 0.93, 0.63, and 2.11 eV, respectively. Furthermore, we have investigated how the diffusion barriers for Li, Na, and K change with different loading levels inside the same pore. Figure 8 clearly shows that the diffusion barriers decrease with increasing loading. Namely, **(a)** in the case of Li@BN:$V_B$, when the pore is partially occupied by one Li atom, the barrier for the second Li atom reduces to 0.79 eV. It further decreases when the loading is increased to two Li atoms, with the third Li atom requiring only 0.47 eV to move to the second most stable position. **(b)** For Na@BN:$V_B$, a more rapid diffusion is observed as compared to Li. The first Na needs to overcome a diffusion barrier of 0.63 eV, which is reduced to 0.31 eV when another Na atom is placed in the pore, increasing the likelihood of hopping to another position. The diffusion barrier of Na atoms continues to decrease with increasing loading concentration, reaching a very low value of 0.08 eV for the third Na atom to diffuse when the pore is loaded with two Na atoms. **(c)** Lastly, for the case of K@BN:$V_B$, K shows the weakest diffusion characteristics compared to the Li and Na. This can be explained by the strong bonding exhibited by the K atom, as shown in the ELF plot in Figure 6. The high diffusion barrier of 2.11 eV for a single K atom is reduced to 0.60 eV for the fourth K atom when the pore is occupied by three K atoms.

Our findings indicate that increasing the loading of MA in the pore significantly lowers the diffusion barriers for Li, Na, and K. This behavior suggests that higher concentrations of MA could facilitate faster atom mobility within the battery material, potentially improving the charge and discharge rates. Na atoms in particular show the most favorable diffusion characteristics, making BN:$V_B$ a promising anode material for NIBs.

### 3.4 $H_2$ storage using 4MA@BN:$V_B$

The other important aim of this work is to evaluate the potential of BN:$V_B$ as efficient $H_2$ storage material. First, the adsorption of a single $H_2$ molecule on Pr-BN and BN:$V_B$ is evaluated without MA decoration. We have carried out relaxation of a single $H_2$ molecule at various adsorption sites on the primitive cell of the two systems. All the adsorption processes lead to physisorption, with maximum $E_{ads}$ and minimum adsorption distance shown in Table 3. The maximum $E_{ads}$ are -0.15 and -0.16 eV for Pr-BN and BN:$V_B$, with a minimum distance of 2.97 Å and 2.57 Å, respectively. Accordingly, we used 4MA@BN:$V_B$ to enhance the $H_2$ adsorption. The goal is to achieve high $H_2$ $E_{ads}$ within the range of range [-0.15, -0.60] eV set by the US-DOE. The MA on BN:$V_B$ are expected to introduce changes to the lattice bonding and contribute to forming electric dipole moments that make the substrate even more polar. The enhancement in the surface polarity could induce dipole moments to the $H_2$ molecules.

#### 3.4.1 Hydrogenation of 4MA@BN:$V_B$

Subsequently, we have evaluated the $H_2$ storage capacities of the 4MA@BN:$V_B$. The $H_2$ adsorption is carried out in a stepwise manner, starting with introducing a single $H_2$ molecule to each MA ($4H_2$ on

4MA@BN:$V_B$ per primitive cell) and optimizing the systems. Next, the number of $H_2$ molecules are increased one by one up to six per MA on 4MA@BN:$V_B$, and structural relaxations are performed in each case. Our results suggest that 4Li-, 4Na-, and 4K@BN:$V_B$ can adsorb up to 20, 24, and 24$H_2$ molecules, respectively. Figure 9 presents the optimized atomic structures with the maximum number of $H_2$ molecules adsorbed on the 4MA@BN:$V_B$ systems. The $E_{ads}$ values per $H_2$ are calculated using equation (4). In Figure 10(a), the absolute values of the corresponding average $H_2$ adsorption energies are presented, demonstrating improvement in $E_{ads}$ compared to the case of BN:$V_B$ with -0.16 eV for a single $H_2$ molecule. These $H_2$ molecules on 4MA@BN:$V_B$ are physiosorbed *via* the predominant electrostatic and van der Waals-type interactions, and the average $E_{ads}$ values are -0.198, -0.215, and -0.192 eV for a maximum hydrogenation (i.e., 20$H_2$, 24$H_2$, and 24$H_2$ adsorbed on 4Li-, 4Na-, and 4K@BN:$V_B$, respectively).

Next, the theoretical $H_2$ gravimetric capacities ($C_T$) are calculated based on the maximum number of $H_2$ molecules ($N_T$), according to equation (5) with data shown in Table 4. All atomic weights are sourced from the database [70]. The calculated gravimetric capacities are 10.64, 10.72, and 9.38 wt%, corresponding to 4Li-, 4Na-, and 4K@BN:$V_B$, respectively, as shown in Figure 10(b). Since DFT utilizes the Born-Oppenheimer frozen lattice approximation and runs all calculations at 0 K, these theoretical capacities exclude the essential thermodynamic contributions of finite pressure and temperature. The adjusted values under the ambient conditions should be calculated using the thermodynamic analysis which will be discussed next.

#### 3.4.2 Thermodynamic analysis

It is important to realize the effective $H_2$ storage capacity of the proposed materials at ambient conditions of temperature (T) and pressure (P). Therefore, we performed statistical thermodynamic analysis based on the Langmuir adsorption model [71-73]. Based on equations (6-8), Figure 11 shows the number of $H_2$ molecules (N) adsorbed on 4MA@ BN:$V_B$ at given P and T values. The N-P-T analysis indicates that $H_2$ molecules can be stored in the storage material at low-T and high-P conditions, then released under high-T and low-P conditions. We further assessed the realistic absorption-desorption characteristics of the adsorbed $H_2$ molecules. In practice, $H_2$ molecules are adsorbed at P of 30 atm and T of 25 °C, and desorbed at 3 atm and 100 °C [74]. Accordingly, Table 4 lists the calculated theoretical storage at 0 K ($C_T$) and effective storage ($C_E$) based on the practical adsorption and desorption conditions. The obtained effective $H_2$ gravimetric capacities of 4Li-, 4Na-, and 4K@BN:$V_B$ are 9.22, 9.89, and 6.34 wt%, respectively. Remarkably, the effective capacities of 4MA@ BN:$V_B$ surpass the target of 5.5 wt% [75].

Table 5 presents a reasonable comparison between our results and those from similar DFT studies on different 2D materials proposed for $H_2$ storage. The computational methods used in each study are highlighted. Overall, our theoretical gravimetric capacities of 4MA@BN:$V_B$ are higher than most of the reported materials. It should be emphasized that in the few studies with higher gravimetric capacities, the authors carried out the hydrogenation to much lower thresholds of average adsorption energies. Thus, the 4MA@BN:$V_B$ systems can surpass their counterparts by accommodating more $H_2$ molecules if the threshold of average $E_{ads}$ reduced than the currently considered (i.e., $|E_{ads}^{crit}|$< 0.16 eV). Based on our findings, it is thus conclusive that 4MA@ BN:$V_B$ are promising materials for $H_2$ storage applications.

## 4. **Conclusion**

In this study, we reported vacancy induced boron nitride monolayers (BN:$V_B$) as multifunctional materials, which could be used as anodes for MIBs and $H_2$ storage applications. Based on comprehensive theoretical approaches, such as DFT, AIMD, and thermodynamic analysis, we studied the structural, electronic, and MA decorations on the surface of BN:$V_B$. Our findings revealed that the presence of a pore in BN:$V_B$ induces robust ionic bonds with MA. The thermal stabilities of 4MA@BN:$V_B$ (MA= Li, Na, K) systems are confirmed through AIMD at 400K. The electronic properties calculation through spin-polarized PDOS, band structure, and Bader charge analysis show significant charge transfer from the MA to BN:$V_B$, improving the electrical conductivity. For MIBs anodes, important parameters of MA@BN:$V_B$, such as specific capacities, electronic characteristics, diffusion barriers, and open-circuit voltage (OCVs), are explored. The theoretical specific capacities of 1821.53, 786.11 and 490.51 mA $hg^{-1}$ are calculated for Li, Na, and K, respectively. Strong electrochemical stabilities are validated by average OCVs of 0.15, 0.25, and 0.32 V for Li, Na, and K, respectively. Diffusion analysis showed lower barriers with increased atom loading, especially in the Na system. For $H_2$ storage applications, 4Li-, 4Na-, and 4K@BN:$V_B$ adsorbed multiple $H_2$ molecules, reaching to significantly gravimetric densities of 10.64, 10.72, and 9.38 wt%, respectively. These values comfortably exceeded the US-DOE target of 5.5 wt%. Average adsorption energies per $H_2$ on the studied systems fall within the ideal range for ambient applications. Further, thermodynamic analysis based on Langmuir adsorption model was employed to study the $H_2$ adsorption/desorption at practical working conditions of fuel cell. Ultimately, MA@BN:$V_B$ serves as a promising multifunctional 2D materials for energy storage applications, especially anode components of MIBs and high reversible capacity in material-based $H_2$ storage.

## CRediT authorship contribution statement

**Wadha Al-Falasi:** Investigation, Formal analysis, Methodology, Visualization, Validation, Project administration, Software, Writing - original draft preparation. **Wael Othman:** Investigation, Validation, Formal analysis, Writing - original draft preparation. **Tanveer Hussain:** Conceptualization, Investigation, Formal analysis, Writing- Reviewing and Editing. **Nacir Tit:** Conceptualization, Funding acquisition, Resources, Supervision, Formal analysis, Writing- Reviewing and Editing.

## Declaration of competing interest

The authors declare that they have no known competing financial interests or personal relationships that could have appeared to influence the work reported in this paper.

## Acknowledgment

This work was supported by the National Water and Energy Center at the United Arab Emirates University under research funding (Grant No. 12R-162).

## References

[1] N. Nitta, F. Wu, J. T. Lee, G. Yushin, Li-ion battery materials: present and future, Mater. Today 18 (2015) 252–264.

[2] K. Xu, Electrolytes and Interphases in Li-Ion Batteries and Beyond, Chem. Rev. 114 (2014) 11503–11618.

[3] N. A. Kaskhedikar, J. Maier, Lithium Storage in Carbon Nanostructures, Adv. Mater. 21 (2009) 2664–2680.

[4] A.P. Yuda, P.Y.E. Koraag, F. Iskandar, H.S. Wasisto, A. Sumboja, Advances of the top-down synthesis approach for high-performance silicon anodes in Li-ion batteries, J. Mater. Chem. A 9 (2021) 18906–18926.

[5] T. Insinna, E.N. Bassey, K. Märker, A. Collauto, A.L. Barra, C.P. Grey, Graphite Anodes for Li-Ion Batteries: An Electron Paramagnetic Resonance Investigation, Chem. Mater. 35 (2023) 5497–5511.

[6] Y. Guo, D. Liu, B. Huang, L. Wang, Q. Xia, A. Zhou, Effects of surface compositions and interlayer distance on electrochemical performance of $Mo_2CT_x$ MXene as anode of Li-ion batteries, J. Phys. Chem. Solids. 176 (2023) 111238.

[7] S. Khammuang, A. Pratumma, A. Sakulkalavek, T. Kaewmaraya, T. Hussain, K. Kotmool, First-principles study of 2H-$Mo_2C$-based MXenes under biaxial strain as Li-battery anodes, Phys. Chem. Chem. Phys. 25 (2023) 19612–19619.

[8] B. Ellis, K. T. Lee, L. Nazar, Positive Electrode Materials for Li-Ion and Li-Batteries, Chem. Mater. 22 (2010) 691–714.

[9] N. Yabuuchi, K. Kubota, M. Dahbi, S. Komaba, Research Development on Sodium-Ion Batteries, Chem. Rev. 114 (2014) 11636–11682.

[10] X. Tan, J. Guo, H. Wang, Z. Qiu, Q. Wang, H. Shu, Pristine and defective 2D SiCN substrates as anode materials for sodium-ion batteries, J. Energy Storage 93 (2024) 112331.

[11] Y. Liang, H. Dong, D. Aurbach, Y. Yao, Current status and future directions of multivalent metal-ion batteries, Nat. Energy 5 (2020) 646–656.

[12] M.S. Kiai, O. Eroglu, N. Aslfattahi, Metal-Ion Batteries: Achievements, Challenges, and Prospects, Crystals 13 (2023) 1002.

[13] M.J. Saadh et al., Performances of nanotubes and nanocages as anodes in Na-ion battery, K-ion battery, and Mg-ion battery, Ionics 30 (2024) 2657–2664.

[14] Y. Li, Y.F. Guo, Z.X. Li, P.F. Wang, Y. Xie, T.F. Yi, Carbon-based nanomaterials for stabilizing zinc metal anodes towards high-performance aqueous zinc-ion batteries, Energy Storage Mater. 67 (2024) 103300.

[15] M. Tang, C. Wang, U. Schwingenschlögl, G. Yang, BC6P Monolayer: Isostructural and Isoelectronic Analogues of Graphene with Desirable Properties for K-Ion Batteries, Chem. Mater. 33 (2021) 9262–9269.

[16] H. Bao et al., Anti-Freezing Electrolytes in Aqueous Multivalent Metal-Ion Batteries: Progress, Challenges, and Optimization Strategies, Chem. Rec. 24 (2024) e202300212.

[17] K. Shi et al., Recent Progress and Prospects on Sodium-Ion Battery and All-Solid-State Sodium Battery: A Promising Choice of Future Batteries for Energy Storage, Energy Fuels 38 (2024) 9280–9319.

[18] K. Kumar, R. Kundu, Empowering Energy Storage Technology: Recent Breakthroughs and Advancement in Sodium-Ion Batteries, ACS Appl. Energy Mater. 7 (2024) 3523–3539.

[19] Z. Bai et al., Low-Temperature Sodium-Ion Batteries: Challenges and Progress, Adv. Energy Mater. 14 (2024) 2303788.

[20] C. Yu et al., Core-shell engineering of titanium-based anodes toward enhanced electrochemical lithium/sodium storage performance: a review, Mater. Today Energy 43 (2024) 101589.

[21] S. Ma et al., Recent advances in carbon-based anodes for high-performance sodium-ion batteries: Mechanism, modification and characterizations, Mater. Today 75 (2024) 334–358.

[22] Z. Zhao et al., Passivation Layers in Mg-Metal Batteries: Robust Interphases for Li-Metal Batteries, Adv. Mater. 36 (2024) 2402626.

[23] T. Wen et al., Interfacial chemistry of anode/electrolyte interface for rechargeable magnesium batteries, J. Magnes. Alloys, Apr. 2024, doi: 10.1016/j.jma.2024.04.010.

[24] Y. Wu et al., Deciphering Unexplored Reversible Potassium Storage and Small Volume Change in a $CaV_4O_9$ Anode within Situ Transmission Electron Microscopy, Adv. Funct. Mater. 34 (2024) 2314344.

[25] B. Vedhanarayanan, K. C. Seetha Lakshmi, Beyond lithium-ion: emerging frontiers in next-generation battery technologies, Front. Batter. Electrochem. 3 (2024) 1377192.

[26] 'Hydrogen Storage', Energy.gov. Accessed: Jun. 05, 2024. [Online]. Available: https://www.energy.gov/eere/fuelcells/hydrogen-storage

[27] 'DOE Technical Targets for Onboard Hydrogen Storage for Light-Duty Vehicles', Energy.gov. Accessed: Jun. 05, 2024. [Online]. Available: https://www.energy.gov/eere/fuelcells/doe-technical-targets-onboard-hydrogen-storage-light-duty-vehicles

[28] M.P. Suh, H.J. Park, T.K. Prasad, D.W. Lim, Hydrogen Storage in Metal–Organic Frameworks, Chem. Rev. 112 (2012) 782–835.

[29] J. Sculley, D. Yuan, H.C. Zhou, The current status of hydrogen storage in metal–organic frameworks—updated, Energy Environ. Sci. 4 (2011) 2721–2735.

[30] A.P. Côté, A.I. Benin, N.W. Ockwig, M. O'Keeffe, A.J. Matzger, O.M. Yaghi, Porous, Crystalline, Covalent Organic Frameworks, Science 310 (2005) 1166–1170.

[31] S.S. Han, H. Furukawa, O.M. Yaghi, W.A.I. Goddard, Covalent Organic Frameworks as Exceptional Hydrogen Storage Materials, J. Am. Chem. Soc. 130 (2008) 11580–11581.

[32] K. Cousins, R. Zhang, Highly Porous Organic Polymers for Hydrogen Fuel Storage, Polymers 11 (2019) 11040690.

[33] M. Mushtaq, S. Khan, N. Tit, Magnetization effect of Mn-embedded in $C_2N$ on hydrogen adsorption and gas-sensing properties: Ab-initio analysis, Appl. Surf. Sci. 537 (2021) 147970.

[34] S.P. Kaur, T. Hussain, T. Kaewmaraya, T.J.D. Kumar, Reversible hydrogen storage tendency of light-metal (Li/Na/K) decorated carbon nitride ($C_9N_4$) monolayer, Int. J. Hydrog. Energy 48 (2023) 26301–26313.

[35] 'Home | Hydrogen Program'. Accessed: Jun. 05, 2024. [Online]. Available: https://www.hydrogen.energy.gov/

[36] J. Ian Jason et al., Defects induced metallized boron hydride monolayers as high-performance hydrogen storage architecture, Int. J. Hydrog. Energy 50 (2024) 455–463.

[37] Q. Li et al., Thermodynamics and kinetics of hydriding and dehydriding reactions in Mg-based hydrogen storage materials, J. Magnes. Alloys 9 (2021) 1922–1941.

[38] S. Bosu, N. Rajamohan, Recent advancements in hydrogen storage - Comparative review on methods, operating conditions and challenges, Int. J. Hydrog. Energy 52 (2024) 352–370.

[39] S. Marchesini, X. Wang, C. Petit, Porous Boron Nitride Materials: Influence of Structure, Chemistry and Stability on the Adsorption of Organics, Front. Chem. 7 (2019) 160.

[40] R. Beiranvand, S. Valedbagi, Electronic and optical properties of h-BN nanosheet: A first principles calculation, Diam. Relat. Mater. 58 (2015) 190–195.

[41] K. Watanabe, T. Taniguchi, H. Kanda, Direct-bandgap properties and evidence for ultraviolet lasing of hexagonal boron nitride single crystal, Nat. Mater. 3 (2004) 404–409.

[42] R.J.P. Román et al., Band gap measurements of monolayer h-BN and insights into carbon-related point defects, 2D Mater. 8 (2021) 044001.

[43] M. Kolos, F. Karlický, Accurate many-body calculation of electronic and optical band gap of bulk hexagonal boron nitride, Phys. Chem. Chem. Phys. 21 (2019) 3999–4005.

[44] Y. Kobayashi, C.L. Tsai, T. Akasaka, Optical band gap of h-BN epitaxial film grown on c -plane sapphire substrate, Phys. Status Solidi C 7 (2010) 1906–1908.

[45] A.K. Dąbrowska et al., Defects in layered boron nitride grown by Metal Organic Vapor Phase Epitaxy: luminescence and positron annihilation studies, J. Lumin. 269 (2024) 120486.

[46] X. Liu et al., Constructing two-dimensional holey graphyne with unusual annulative π-extension, Matter 5 (2022) 2306–2318.

[47] Y. Gao, H. Zhang, H. Pan, Q. Li, J. Zhao, Ultrahigh hydrogen storage capacity of holey graphyne, Nanotechnology 32 (2021) 215402.

[48] V. Mahamiya, A. Shukla, N. Garg, B. Chakraborty, High-capacity reversible hydrogen storage in scandium decorated holey graphyne: Theoretical perspectives, Int. J. Hydrog. Energy 47 (2022) 7870–7883.

[49] V. Mahamiya, A. Shukla, B. Chakraborty, Prediction of a Novel 2D Porous Boron Nitride Material with Excellent Electronic, Optical and Catalytic Properties, Phys. Chem. Chem. Phys. 24 (2022) 21009-21019.

[50] J. Hafner, Ab-initio simulations of materials using VASP: Density-functional theory and beyond, J. Comput. Chem. 29 (2008) 2044–2078.

[51] M. Ernzerhof, G.E. Scuseria, Assessment of the Perdew–Burke–Ernzerhof exchange-correlation functional, J. Chem. Phys. 110 (1999) 5029–5036.

[52] S. Grimme, J. Antony, S. Ehrlich, H. Krieg, A consistent and accurate ab initio parametrization of density functional dispersion correction (DFT-D) for the 94 elements H-Pu, J. Chem. Phys. 132 (2010) 154104.

[53] H.J. Monkhorst, J. D. Pack, Special points for Brillouin-zone integrations, Phys. Rev. B 13 (1976) 5188.

[54] G. Henkelman, A. Arnaldsson, and H. Jónsson, A fast and robust algorithm for Bader decomposition of charge density, Comput. Mater. Sci. 36 (2006) 354–360.

[55] Liu, T., Jin, Z., Liu, DX. *et al.* A density functional theory study of high-performance pre-lithiated $MS_2$ (M = Mo, W, V) Monolayers as the Anode Material of Lithium Ion Batteries, Sci. Rep. **10** (2020) 6897.

[56] F. Keshavarz, M. Kadek, B. Barbiellini, A. Bansil, Anodic Activity of Hydrated and Anhydrous Iron (II) Oxalate in Li-Ion Batteries, Condens. Matter 7 (2022) 8.

[57] W. Othman, W. AlFalasi, T. Hussain, N. Tit, Light-Metal Functionalized Boron Monoxide Monolayers as Efficient Hydrogen Storage Material: Insights from DFT Simulations. arXiv, Feb. 22, 2024. doi: 10.48550/arXiv.2402.14342.

[58] A. Hashmi, M.U. Farooq, I. Khan, J. Son, J. Hong, Ultra-high capacity hydrogen storage in a Li decorated two-dimensional $C_2N$ layer, J. Mater. Chem. A 5 (2017) 2821–2828.

[59] N. Mahar, Q. Yuan, L. Cheng, A single Mn atom supported by a boron-vacancy in a BN monolayer: An encouraging catalyst for CO oxidation', Mol. Catalysis 563 (2024) 114260.

[60] H. Li, L. Xue, Z. Wang, H.J. Liu, Point vacancy defects in hexagonal boron nitride studied by first-principles, JJAP Conf. Proc. 9 (2023) 011104–011104.

[61] A.M. Satawara, G.A. Shaikh, S.K. Gupta, P.N. Gajjar, Structural, electronic and optical properties of hexagonal boron-nitride (h-BN) monolayer: An Ab-initio study, Mater. Today Proc. 47 (2021) 529–532.

[62] F. Keshavarz, M. Kadek, B. Barbiellini, A. Bansil, Anodic Activity of Hydrated and Anhydrous Iron (II) Oxalate in Li-Ion Batteries, Condens. Matter 7 (2022) 8.

[63] A. Samad, A. Shafique, U. Schwingenschlögl, Z. Ji, G. Luo, Monolayer, Bilayer, and Bulk BSi as Potential Anode Materials of Li-Ion Batteries, ChemPhysChem 23 (2022) e202200041.

[64] N. Kashhedikar, J. Maier,Lithium Storage in Carbon Nanostructures, Adv. Mater. 21 2664-2680.

[65] C. Chen et al., Two-dimensional $PC_3$ monolayer as promising hosts of Li-ion storage: A first-principles calculations study, Colloids Surf. Physicochem. Eng. Asp. 685 (2024) 133313.

[66] A. Hussain et al., Improving the electrochemical performance of sodium-ion batteries with $Zn_{1-x}Mn_xO$ anode (0≤x≤0.2), Int. J. Hydrog. Energy 71 (2024) 894–902.

[67] H. Liu, Y. Cai, Z. Guo, J. Zhou, Two-Dimensional V2N MXene Monolayer as a High-Capacity Anode Material for Lithium-Ion Batteries and Beyond: First-Principles Calculations, ACS Omega 7 (2022) 17756–17764.

[68] Y. Shen, J. Liu, X. Li, Q. Wang, Two-Dimensional T-NiSe2 as a Promising Anode Material for Potassium-Ion Batteries with Low Average Voltage, High Ionic Conductivity, and Superior Carrier Mobility, ACS Appl. Mater. Interfaces 11 (2019) 35661–35666.

[69] C. Pu, Z. Wang, X. Tang, D. Zhou, J. Cheng, A Novel Two-Dimensional $ZnSiP_2$ Monolayer as an Anode Material for K-Ion Batteries and $NO_2$ Gas Sensing, Molecules 27 (2022) 6726.

[70] T. Prohaska et al., Standard atomic weights of the elements 2021 (IUPAC Technical Report), Pure Appl. Chem. 94 (2022) 573–600.

[71] H. Bae et al., High-throughput screening of metal-porphyrin-like graphenes for selective capture of carbon dioxide, Sci. Rep. 6 (2016) 21788.

[72] H. Yang, H. Bae, M. Park, S. Lee, K.C. Kim, H. Lee, Fe–Porphyrin-like Nanostructures for Selective Ammonia Capture under Humid Conditions, J. Phys. Chem. C122 (2018) 2046–2052.

[73] N. Kumar et al., First-Principles Approach for Assessing the Detection of Alzheimer's Biomarkers Using Titanium Carbide MXenes, ACS Appl. Nano Mater. 7 (2024) 6873–6884.

[74] H. Lee, W.I. Choi, J. Ihm, Combinatorial Search for Optimal Hydrogen-Storage Nanomaterials Based on Polymers, Phys. Rev. Lett. 97 (2006) 056104.

[75] P. Habibi, T.J.H. Vlugt, P. Dey, O.A. Moultos, Reversible Hydrogen Storage in Metal-Decorated Honeycomb Borophene Oxide, ACS Appl. Mater. Interfaces 13 (2021) 43233–43240.

[76] A. Vaidyanathan, P. Mane, V. Wagh, B. Chakraborty, Vanadium-decorated 2D polyaramid material for high-capacity hydrogen storage: Insights from DFT simulations, J. Energy Storage 78 (2024) 109899.

[77] P. Mane, S.P. Kaur, M. Singh, A. Kundu, B. Chakraborty, Superior hydrogen storage capacity of Vanadium decorated biphenylene (Bi+V): A DFT study, Int. J. Hydrog. Energy 48 (2023) 28076–28090.

[78] Y. Liu, L. Ren, Y. He, H.P. Cheng, Titanium-decorated graphene for high-capacity hydrogen storage studied by density functional simulations, J. Phys. Condens. Matter 22 (2010) 445301.

[79] Z. Liu, T. Hussain, A. Karton, S. Er, Empowering hydrogen storage properties of haeckelite monolayers via metal atom functionalization, Appl. Surf. Sci. 556 (2021) 149709.

[80] B.J. Cid, A.N. Sosa, Á. Miranda, L.A. Pérez, F. Salazar, M. Cruz-Irisson, Hydrogen storage on metal decorated pristine siligene and metal decorated boron-doped siligene, Mater. Lett. 293 (2021) 129743.

[81] P. Panigrahi et al., Selective decoration of nitrogenated holey graphene ($C_2N$) with titanium clusters for enhanced hydrogen storage application, Int. J. Hydrog. Energy 46 (2021) 7371–7380.

[82] P. Mane, S.P. Kaur, B. Chakraborty, Enhanced reversible hydrogen storage efficiency of zirconium-decorated biphenylene monolayer: A computational study, Energy Storage 4 (2022) e377.

[83] P.A. Denis, F. Iribarne, Hydrogen storage in doped biphenylene based sheets, Comput. Theor. Chem. 1062 (2015) 30–35.

[84] Z. Liu, S. Liu, and S. Er, Hydrogen storage properties of Li-decorated B2S monolayers: A DFT study, Int. J. Hydrog. Energy 44 (2019) 16803–16810.

[85] T. Wang, Z. Tian, Yttrium-decorated C48B12 as hydrogen storage media: a DFT study, Int. J. Hydrog. Energy 45 (2020) 24895–24901.

[86] M. Dixit, T.A. Maark, S. Pal, Ab initio and periodic DFT investigation of hydrogen storage on light metal-decorated MOF-5, Int. J. Hydrog. Energy 36 (2011) 10816–10827.

[87] A. Kundu, B. Chakraborty, Yttrium doped covalent triazine frameworks as promising reversible hydrogen storage material: DFT investigations, Int. J. Hydrog. Energy 47 (2022) 30567–30579.

[88] B. Chakraborty, P. Mane, A. Vaidyanathan, Hydrogen storage in scandium decorated triazine based g-C3N4: Insights from DFT simulations, Int. J. Hydrog. Energy 47 (2022) 41878–41890.

[89] L. Yuan et al., Hydrogen storage capacity on Ti-decorated porous graphene: First-principles investigation, Appl. Surf. Sci. 434 (2018) 843–849.

[90] P. Mane, A. Vaidyanathan, B. Chakraborty, Graphitic carbon nitride (g-$C_3N_4$) decorated with Yttrium as potential hydrogen storage material: Acumen from quantum simulations, Int. J. Hydrog. Energy 47 (2022) 41898–41910.

[91] S. Dong et al., Construction of transition metal-decorated boron doped twin-graphene for hydrogen storage: A theoretical prediction, Fuel 304 (2021) 121351.

[92] X. Liang, S.P. Ng, N. Ding, C.M. L. Wu, Strain-induced switch for hydrogen storage in cobalt-decorated nitrogen-doped graphene, Appl. Surf. Sci. 473 (2019) 174–181.

[93] L. Yuan et al., First-principles study of V-decorated porous graphene for hydrogen storage, Chem. Phys. Lett. 726 (2019) 57-61.

[94] S. Chu, L. Hu, X. Hu, M. Yang, J. Deng, Titanium-embedded graphene as high-capacity hydrogen-storage media, Int. J. Hydrog. Energy 36 (2011) 12324–12328.

[95] Z. Zhongming, L. Linong, Y. Xiaona, Z. Wangqiang, L. Wei, Potassium-doped PC71BM for hydrogen storage: Photoelectron spectroscopy and first-principles studies, Int. J. Hydrog. Energy 46 (2021) 13061-13069.

[96] E. Eroglu, S. Aydin, M. Şimşek, Effect of boron substitution on hydrogen storage in Ca/DCV graphene: A first-principles study, Int. J. Hydrog. Energy 44 (2019) 27511–27528.

[97] N. Zheng, S. Yang, H. Xu, Z. Lan, Z. Wang, H. Gu, A DFT study of the enhanced hydrogen storage performance of the Li-decorated graphene nanoribbon’, Vacuum 171 (2020) 109011.

[98] Q. Wu, M. Shi, X. Huang, Z. Meng, Y. Wang, Z. Yang, A first-principles study of Li and Na co-decorated T4,4,4-graphyne for hydrogen storage, Int. J. Hydrog. Energy 46 (2021) 8104–8112.

[99] A.N. Sosa, B.J. Cid, Á. Miranda, L.A. Pérez, G.H. Cocoletzi, M. Cruz-Irisson, 'A DFT investigation: High-capacity hydrogen storage in metal-decorated doped germanene', J. Energy Storage 73 (2023) 108913.

[100] Q. Yin, G. Bi, R. Wang, Z. Zhao, K. Ma, High-capacity hydrogen storage in lithium decorated penta-BN2: A first-principles study, J. Power Sources 591 (2024) 233814.

[101] J. Hao et al., An investigation of Li-decorated N-doped penta-graphene for hydrogen storage, Int. J. Hydrog. Energy 46 (2021) 25533–25542.

[102] Q. Yin, G. Bi, R. Wang, Z. Zhao, K. Ma, High-capacity hydrogen storage in Li-decorated defective penta-$BN_2$: A DFT-D2 study, Int. J. Hydrog. Energy 48 (2023) 26288–26300.

[103] Y. Yong, S. Hu, Z. Zhao, R. Gao, H. Cui, Z. Lv, Potential reversible and high-capacity hydrogen storage medium: Li-decorated B3S monolayers, Mater. Today Commun. 29 (2021) 102938.

[104] Y. Zhang, P. Liu, X. Zhu, Li decorated penta-silicene as a high-capacity hydrogen storage material: A density functional theory study, Int. J. Hydrog. Energy 46 (2021) 4188–4200.

[105] L. Bi et al., Density functional theory study on hydrogen storage capacity of metal-embedded penta-octa-graphene, Int. J. Hydrog. Energy 47 (2022) 32552–32564.

[106] A.L. Marcos-Viquez, A. Miranda, M. Cruz-Irisson, and L.A. Pérez, Tin carbide monolayers decorated with alkali metal atoms for hydrogen storage, Int. J. Hydrog. Energy 47 (2021) 41329–41335.

[107] S. Haldar, S. Mukherjee, C.V. Singh, Hydrogen storage in Li, Na and Ca decorated and defective borophene: a first principles study, RSC Adv. 8 (2018) 20748–20757.

[108] N. Khossossi et al., Hydrogen storage characteristics of Li and Na decorated 2D boron phosphide, Sustain. Energy Fuels 4 (2020) 4538–4546.

[109] T. Kaewmaraya et al., Ultrahigh hydrogen storage using metal-decorated defected biphenylene, Appl. Surf. Sci. 629 (2023) 157391.

[110] S. Lu, S. Zhang, X. Hu, Ab Initio Study of High-Capacity Hydrogen Storage in Lithium-Shrouded Honeycomb Borophene Oxide Nanosheet, J. Phys. Chem. C 126 (2022) 20762–20772.

[111] Y. Wei, F. Gao, J. Du, G. Jiang, Hydrogen storage on Li-decorated B4N: a first-principles calculation insight, J. Phys. Appl. Phys.54 (2021) 445501.

# Tables

Table1: Electronic properties of Pr-BN, BN:$V_B$, and metal-functionalized BN:$V_B$. Δq, $E_g$, $E_F$ and M represent the charge transfer, band gap, Fermi energy, and total magnetization, respectively.

| System | $\Delta q\ (e)$ | $E_g\ (eV)$ | $E_F\ (eV)$ | $M\ (\mu_B)$ |
|---|---|---|---|---|
| Pr-BN | NA | 4.21↑↓ | -4.98 | 0.00 |
| BN:$V_B$ | NA | 3.13↑ 1.18↓ | -4.12 | 1.00 |
| 4Li@BN:$V_B$ | 2.07 | 0.17↑ 1.42↓ (1.14)* | -0.97 | 1.00 |
| 4Na@BN:$V_B$ | 2.12 | 0.15↑ 1.23↓ (0.15)* | -1.23 | 1.00 |
| 4K@BN:$V_B$ | 2.412 | 0.17↑ 1.42↓ (0.05)* | -0.97 | 1.00 |
| 44Li@BN:$V_B$ | 6.47 | 0.00 ↑↓ | -0.02 | 0.00 |
| 28Na@BN:$V_B$ | 2.31 | 0.00 ↑↓ | 0.51 | 0.00 |
| 20K@BN:$V_B$ | 2.86 | 0.00 ↑↓ | -0.21 | 0.00 |

*non-polarized band gap energy

Table2: Theoretical specific capacities ($C_Q$) and open-circuit voltage (OCV) of BN:$V_B$ as an anode material for Li-, Na-, and K-ion batteries.

| **System** | **$C_Q$ ($mAhg^{-1}$)** | **OCV (V)** | **Reference** |
|---|---|---|---|
| Li@BN | 1821.53 | 2.22 | This work |
| Na@BN | 786.11 | 2.82 | This work |
| K@BN | 490.51 | 2.04 | This work |
| Li@2D $PC_3$ | 1200.0 | 0.29 | [66] |
| Li@graphite | 372.0 | - | [3] |
| K@$BC_6P$ | 1410.0 | 0.35 | [10] |
| K@T-$NiSe_2$ | 247.0 | 0.49 | [69] |
| K@$ZnSiP_2$ | 517.0 | - | [70] |
| Na@$Zn_{0.8}Mn_{0.2}O$ | 312.4 | - | [67] |
| Na@p-SiCN | 1486 | 0.34 | [11] |

Table 3: The adsorption energy ($E_{ads}$) and minimum distance (D) of one $H_2$ molecule on Pr-BN and BN:$V_B$

| Structure | $E_{ads}$ (eV) | D (Å) |
|---|---|---|
| **Pr-BN** | -0.15 | 2.97 |
| **BN:$V_B$** | -0.16 | 2.57 |

Table 4: The optimal theoretical numbers of $H_2$ molecules ($N_T$) absorbed on 4Li-, 4Na-, and 4K@BN:$V_B$. These numbers are obtained from the DFT calculations and are used to compute the theoretical storage capacities ($C_T$). Meanwhile, $N_a$ and $N_d$ represent the amount of adsorbed $H_2$ molecules under the adsorption (P = 30 atm and T = 25 °C) and desorption (3 atm and T = 100 °C) conditions, respectively. The practical number of $H_2$ molecules ($N_p$) that can be reversibly stored/released is given by $N_a$-$N_d$. The effective storage capacity ($C_E$) is then computed based on $N_p$.

| **System** | **$C_T$ (wt%)** | **$N_T$** | **$N_a$** | **$N_d$** | **$N_P$** | **$C_E$ (wt%)** |
|---|---|---|---|---|---|---|
| 4Li@BN:$V_B$ | 10.64 | 20 | 18.973 | 1.633 | 17.341 | 9.22 |
| 4Na@BN:$V_B$ | 10.72 | 24 | 23.749 | 6.334 | 17.416 | 9.89 |
| 4K@BN:$V_B$ | 9.38 | 24 | 23.854 | 7.630 | 16.223 | 6.34 |

Table 5: Comparison with different 2D systems, highlighting the theoretical number of adsorbed $H_2$ molecules per simulation cell ($N_T$), average adsorption energy per $H_2$ molecule ($E_{ads}$), theoretical $H_2$ gravimetric capacity ($C_T$), and computational methods used.

| **Hydrogen Storage System** | **$N_T$** | **$E_{ads}$ (eV)** | **$C_T$ (wt%)** | **Reference** |
|---|---|---|---|---|
| Li-decorated BN:$V_B$* | 20 | -0.20 | 10.64 | This work |
| Na-decorated BN:$V_B$* | 24 | -0.22 | 10.72 | This work |
| K-decorated BN:$V_B$* | 24 | -0.19 | 9.38 | This work |
| V-decorated 2DPA-I** | 7 | -0.44 | 7.29 | [76] |
| V-decorated biphenylene* | 7 | -0.47 | 10.30 | [77] |
| Ti-decorated graphene** | 8 | -0.42 | 7.80 | [78] |
| Li-decorated B@$r_{57}$haeckelite* | 12 | -0.16 | 10.00 | [79] |
| Sc-doped Holey graphyne*** | 5 | -0.36 | 9.80 | [48] |
| Li-decorated B-doped siligene* | 4 | -0.17 | 12.71 | [80] |
| Ti-decorated $C_2N$* | 10 | -0.28 | 6.80 | [81] |
| Zr-decorated biphenylene* | 9 | -0.40 | 9.95 | [82] |
| Li-decorated biphenylene***** | 12 | -0.20 | 7.40 | [83] |
| Li-decorated $B_2S$ honeycomb* | 12 | -0.14 | 9.10 | [84] |
| Y-decorated $C_{48}B_{12}$** | 72 | -0.46 | 7.51 | [85] |
| Li-decorated MOF-5** | 18 | _ | 4.30 | [86] |
| Y-decorated covalent triazine frameworks* | 7 | -0.33 | 7.30 | [87] |
| Sc-decorated g-$C_3N_4$** | 7 | -0.39 | 8.55 | [88] |
| Ti-decorated graphene**** | 8 | -0.46 | 6.11 | [89] |
| Y-decorated g-$C_3N_4$** | 9 | -0.33 | 8.55 | [90] |
| Ti-decorated B-doped twin-graphene* | 8 | -0.20 | 4.95 | [91] |
| Co-decorated N-doped graphene** | 28 | -0.19 | 11.36 | [92] |
| V-decorated porous graphene**** | 6 | -0.56 | 4.58 | [93] |
| Ti-decorated graphene** | 8 | -0.21 | 6.30 | [94] |
| K-doped $PC_{71}BM$** | 45 | -0.14 | 6.22 | [95] |
| Ca-decorated DCV graphene**** | 14 | -0.10 | 5.80 | [96] |
| Li-decorated graphene nanoribbons** | 8 | -0.24 | 3.80 | [97] |
| Li-decorated $T_{4,4,4}$-graphyne* | 16 | -0.20 | 10.46 | [98] |
| K-decorated Ga-doped germanene[**] | 36 | -0.19 | 8.19 | [99] |
| Li-decorated P-$BN_2$[**] | 28 | -0.16 | 13.27 | [100] |
| Li-decorated N-doped penta-graphene[**] | 12 | -0.24 | 7.88 | [101] |
| Li-decorated defective penta-$BN_2$[**] | 16 | −0.14 | 9.17 | [102] |
| Li-decorated $B_3S$[*] | 12 | −0.17 | 7.70 | [103] |
| Li-decorated penta-silicene[****] | 12 | -0.22 | 6.42 | [104] |
| Li-decorated penta-octa-graphene[*] | 3 | -0.22 | 9.90 | [105] |
| K-decorated SnC[**] | 6 | -0.20 | 5.50 | [106] |
| Li-decorated borophene[****] | 10 | -0.36 | 9.00 | [107] |
| Li-decorated boron phosphide[**] | 16 | -0.19 | 7.40 | [108] |
| Li-decorated defected biphenylene[*] | 14 | -0.20 | 8.75 | [109] |
| Li-decorated honeycomb borophene oxide[**] | 32 | -0.24 | 8.30 | [75] |
| Li-decorated honeycomb borophene oxide[**] | 8 | -0.22 | 9.84 | [110] |
| Li-decorated $B_4N$[**] | 16 | -0.16 | 6.23 | [111] |

[*]GGA-PBE/DFT-D3
[**]GGA-PBE/DFT-D2
[***]GGA-PBE/DFT-D2 and HSE06
[****]GGA-PBE and DFT-D
[*****]VDW-DF/DZP

# Figures

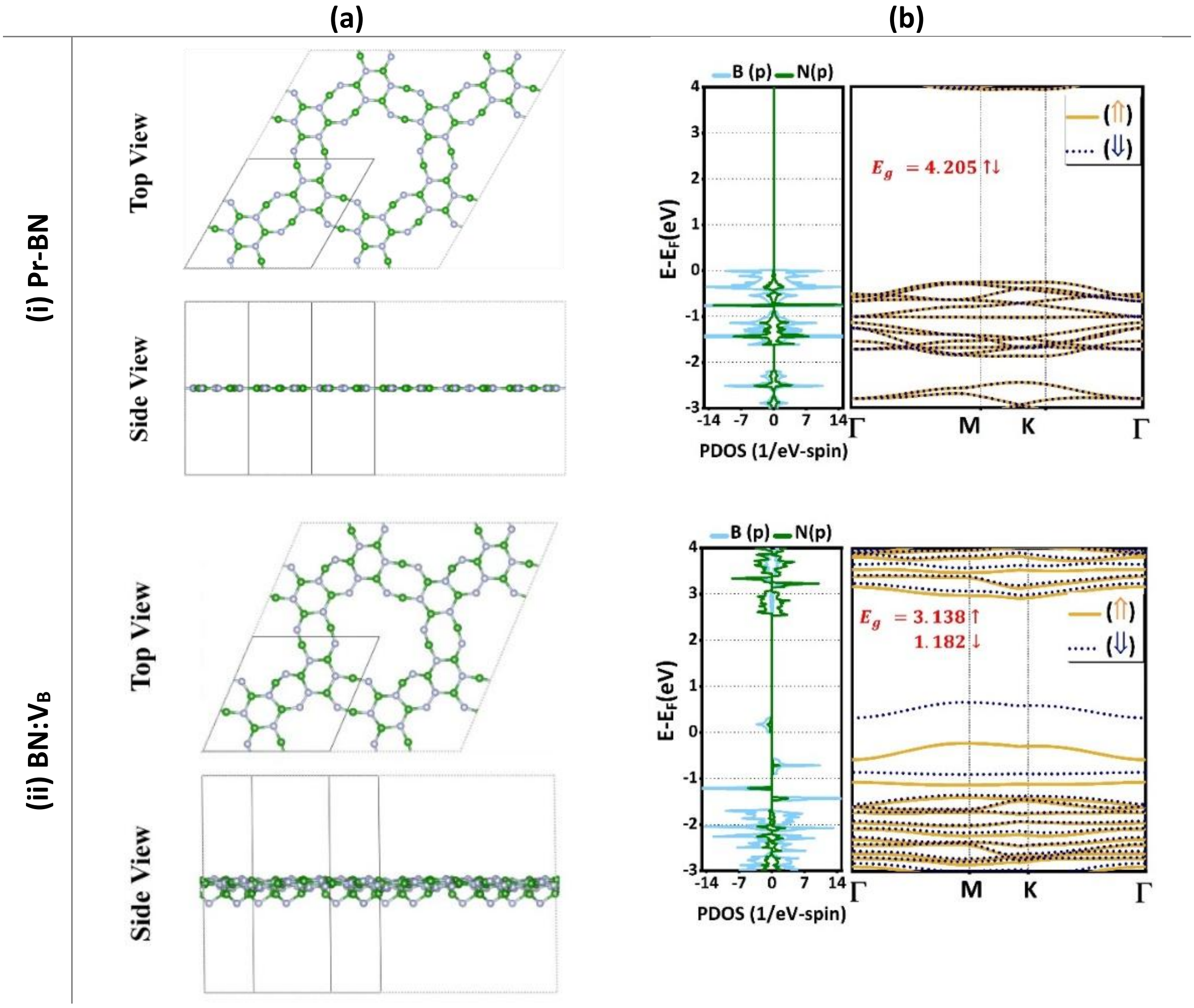

Figure 1: (a) Optimized atomic structures (top and side views) and (b) electronic properties of (i) pristine Boron-Nitride Monolayer (Pr-BN) and (ii) Boron-Nitride Monolayer with Boron vacancy (BN:$V_B$). The primitive cell is represented by the connected lines. Fermi level is taken as an energy reference ($E_F$ = 0) in the spin-polarized band structure and projected density of states plots. Colours of atoms: B (green) and N (light blue).

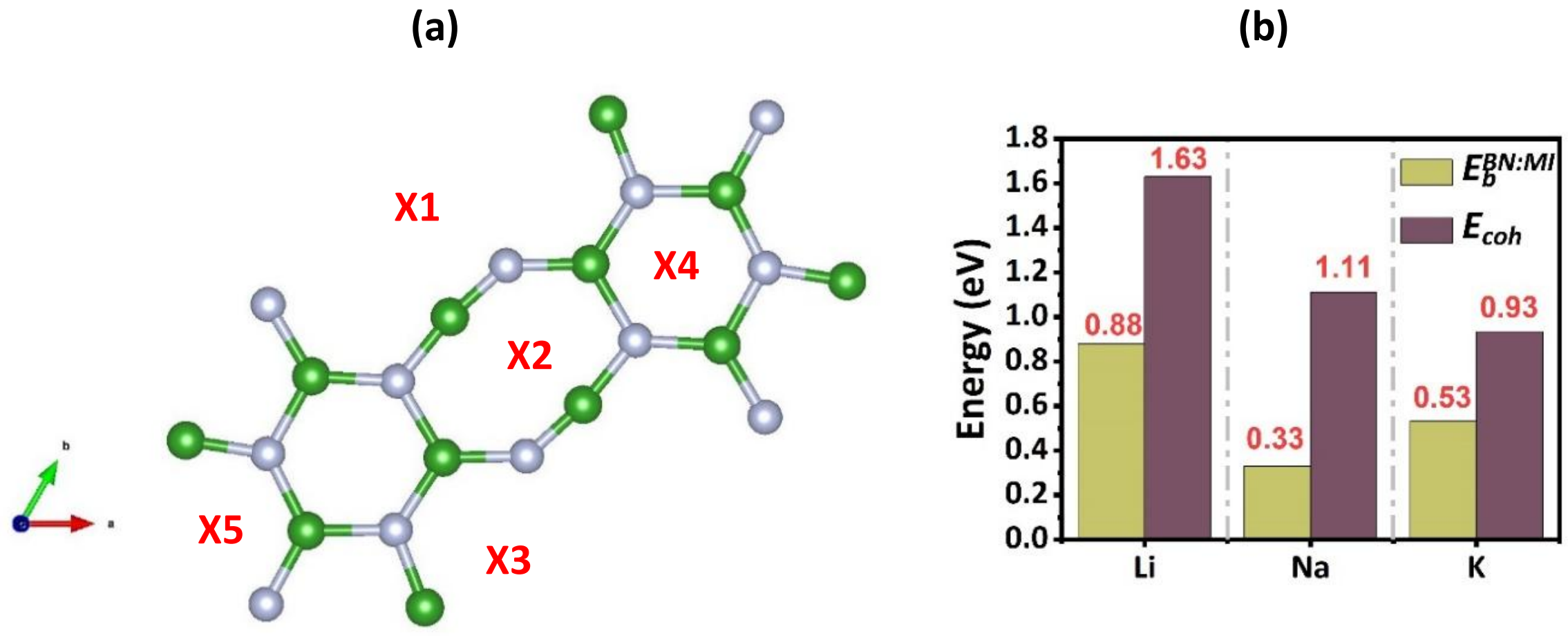

Figure 2: (a) Five sites for the adsorption of MA (Li, K, Na) on Pr-BN. (b) Absolute value of the binding energy of MA are compared to the cohesive energy (shaded brown).

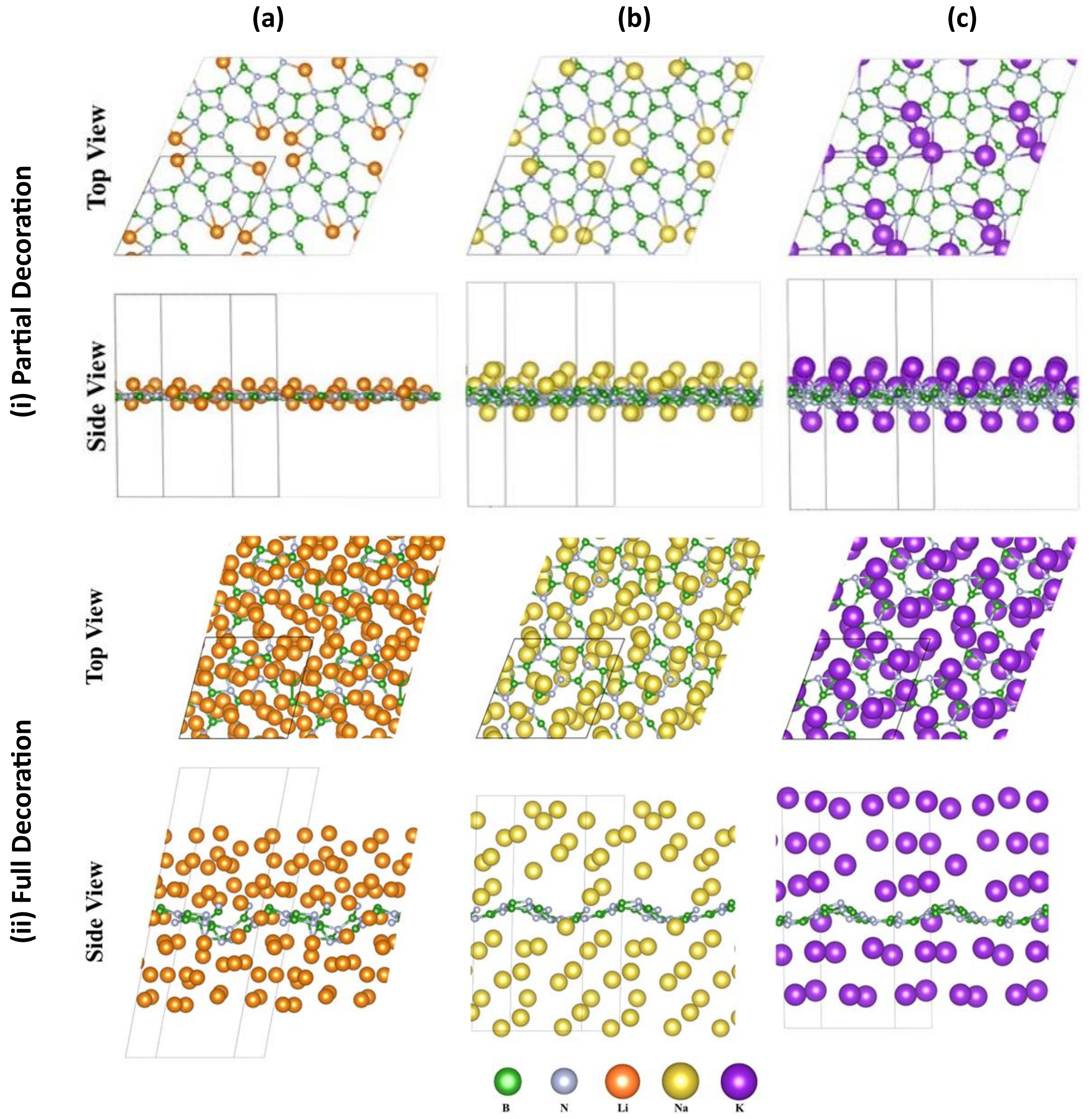

Figure 3: Optimized structures (top and side views) of BN:$V_B$ decorated with (a) Li, (b) Na, and (c) K per primitive cell in (i) partial decoration (i.e., 4MA) and (ii) full decoration (i.e., 44Li, 28 Na, and 20 K)

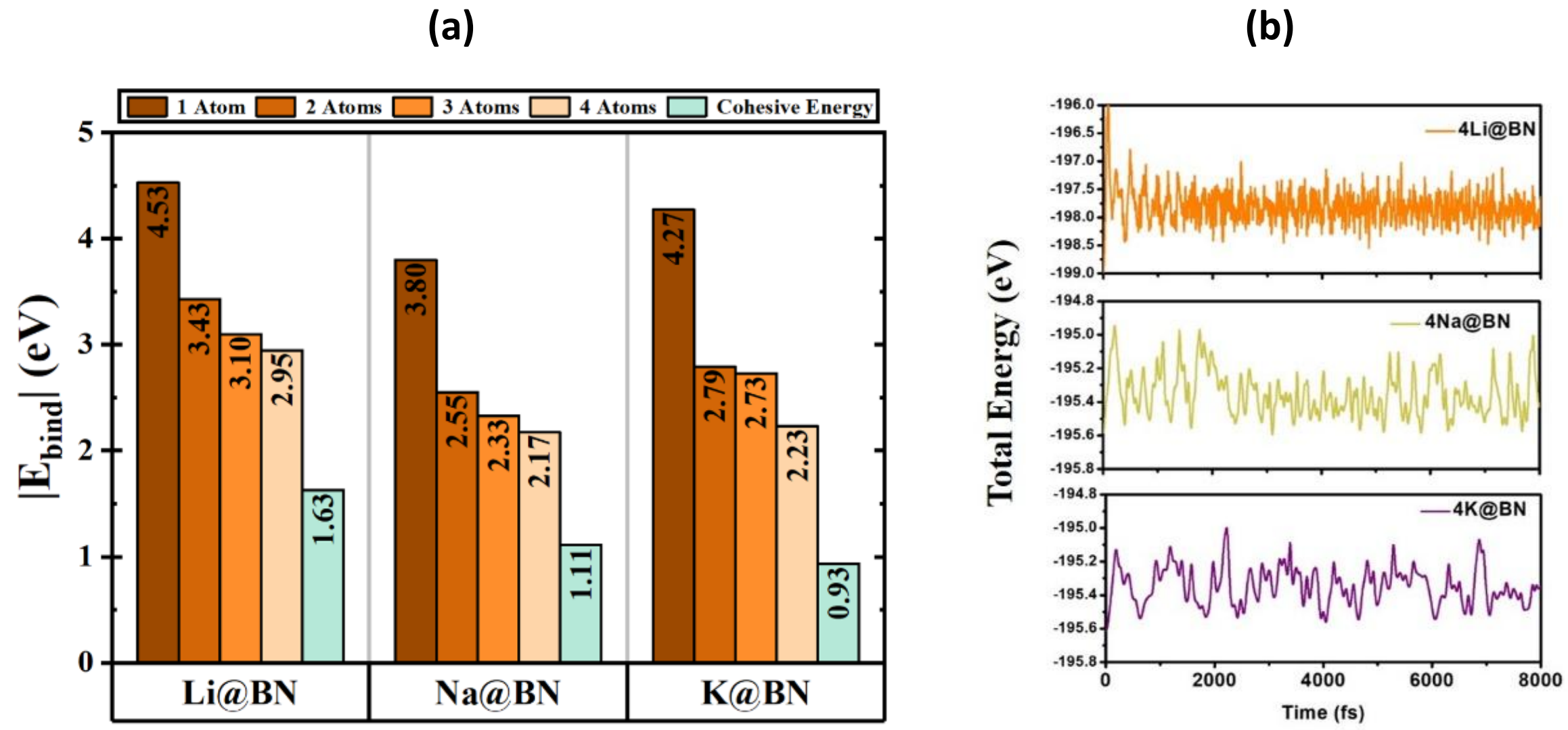

Figure 4: (a) Average binding energy per MA (compared to its cohesive energy) of BN:$V_B$ decorated with 4Li, 4Na, and 4K. (b) Ab-initio molecular dynamics (AIMD) simulation plots of 4MA@BN:$V_B$ at 400 K for a duration of 8 ps.

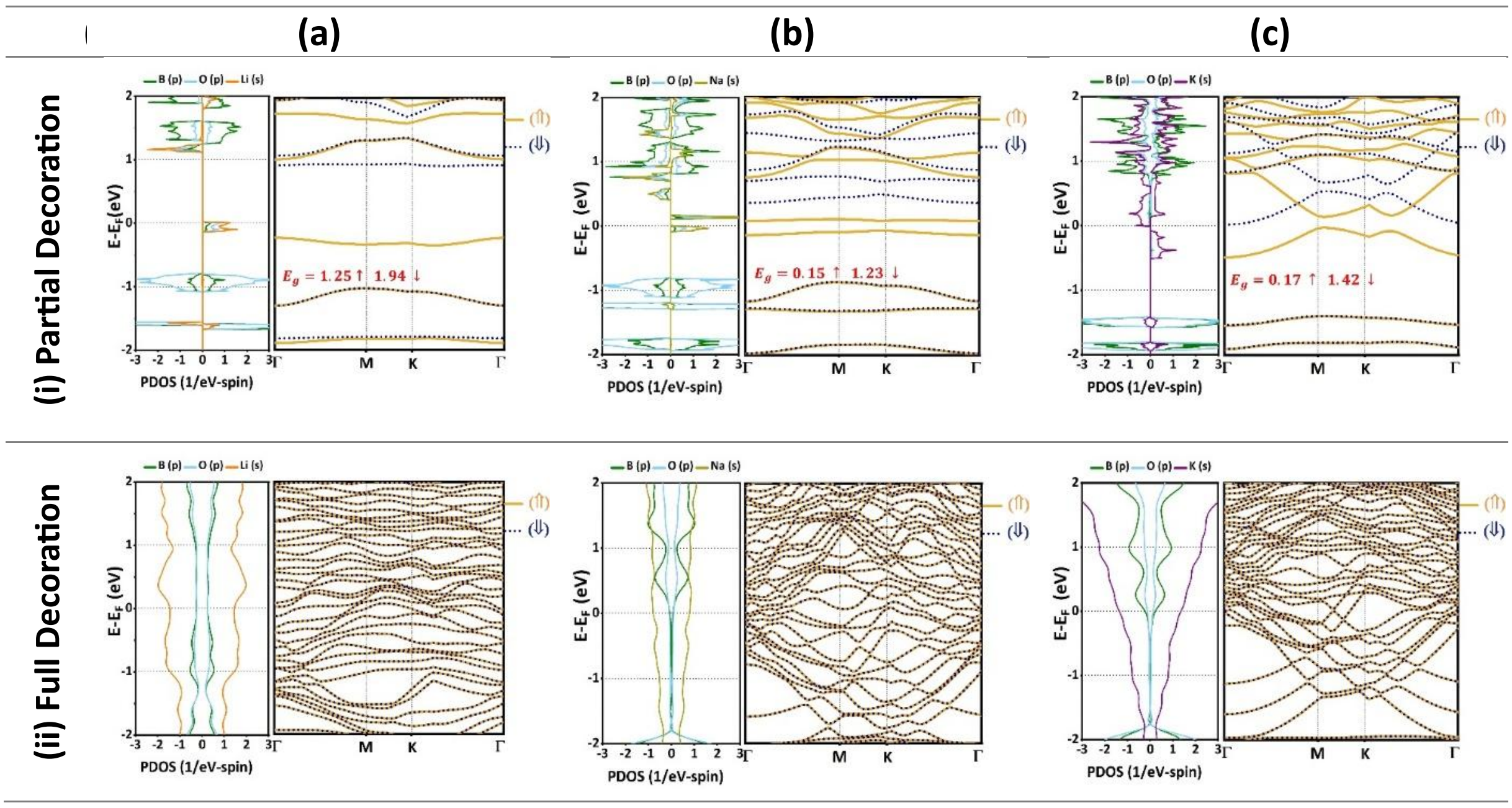

Figure 5: Spin-polarized band structures and projected density of states (PDOS) of BN:$V_B$ decorated with (a) Li, (b) Na, and (c) K in (i) partial decoration (i.e., 4MA) and (ii) full decoration (i.e., 44Li, 28 Na, and 20 K). Fermi level is taken as an energy reference ($E_F$ = 0).

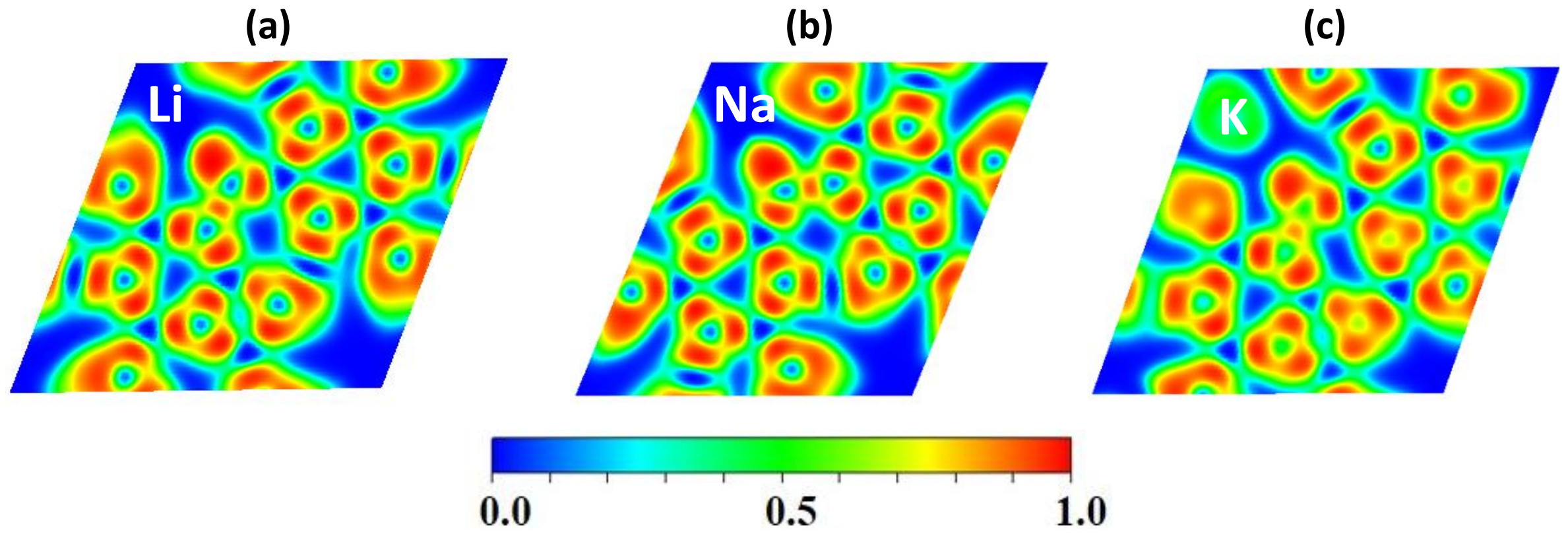

Figure 6: Calculated electron localization function (ELF) maps for BN:$V_B$ decorated with a single (a) Li, (b) Na, and (c) K.

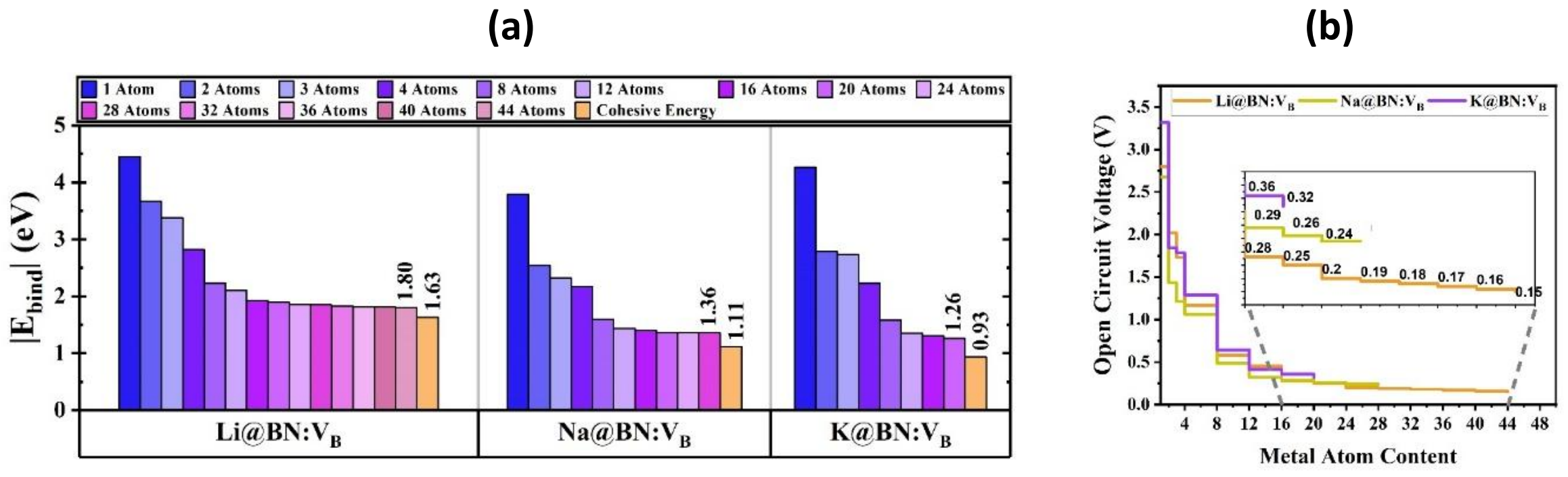

Figure 7: (a) The average binding energy per metal atom for $n$MA@BN:$V_B$. (b) The average open-circuit voltage (OCV) changes with respect to the metal atom content, with a maximum of 44Li, 28Na, and 20K.

(a) (b) (c)

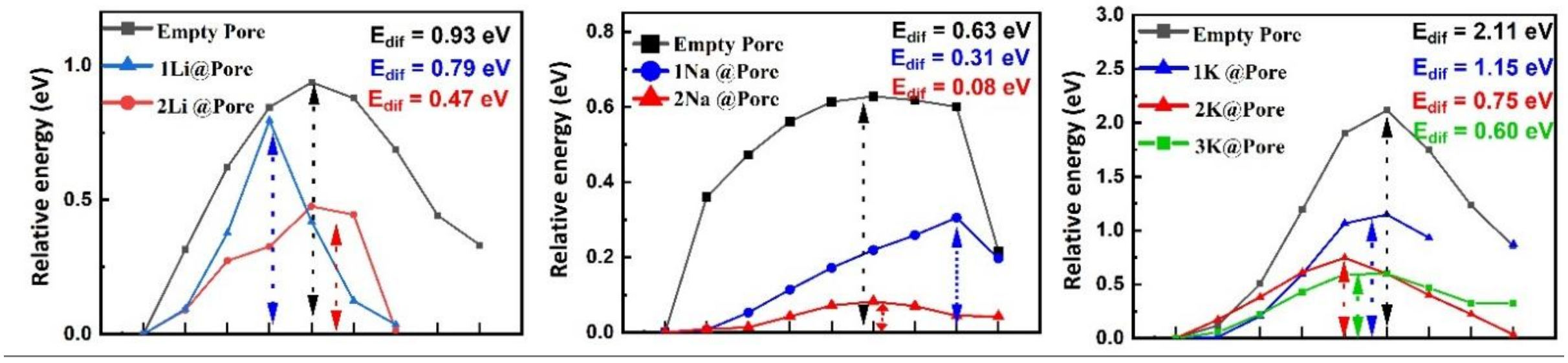

Figure 8: Minimum energy pathways for (a) Li, (b) Na, and (c) K atoms to diffuse on top of BN:$V_B$ starting from the most preferred site with different MA loadings in the pore (*n*MA@BN:$V_B$).

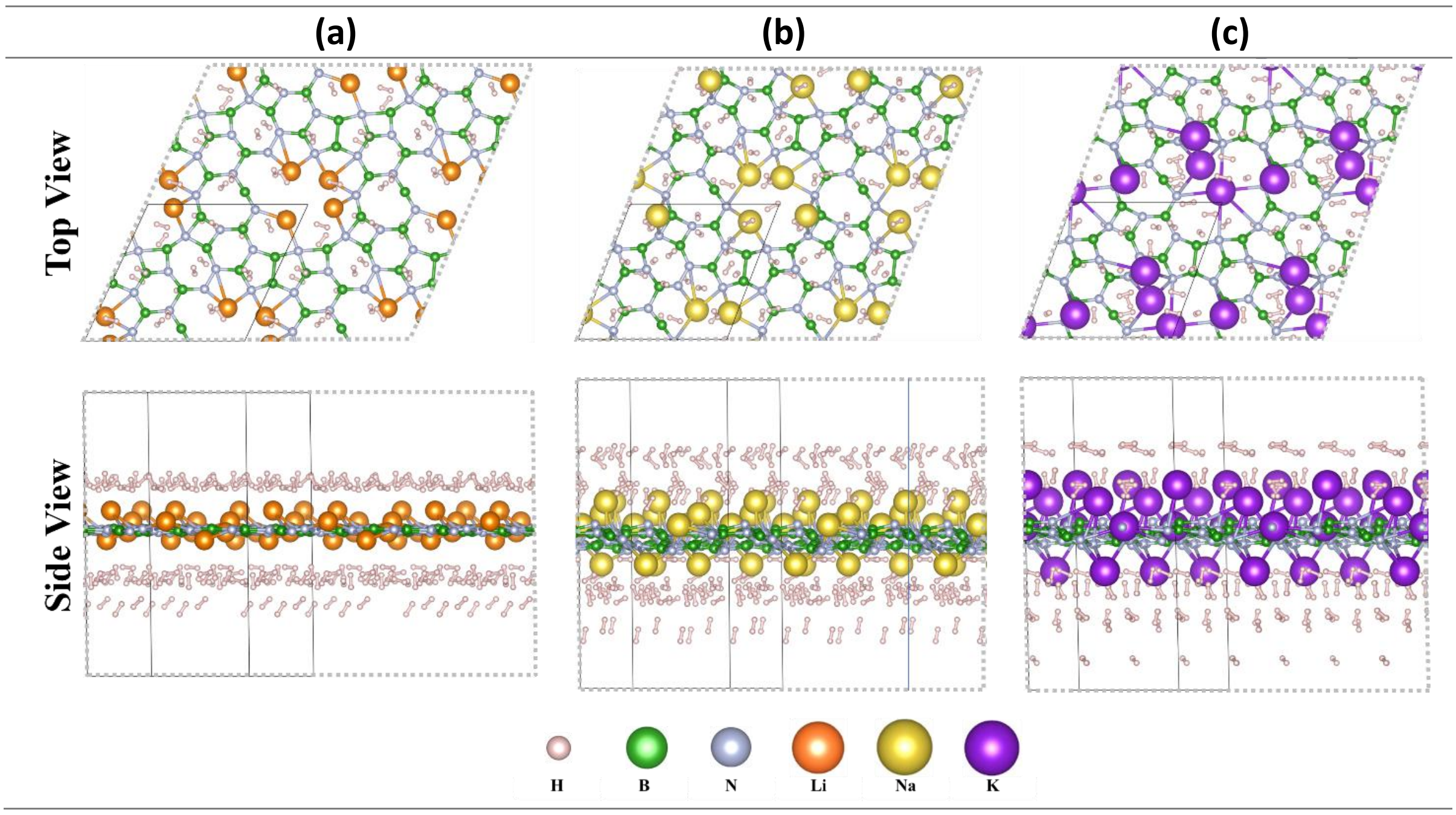

Figure 9: Optimized atomic structures (top and side views) of the maximum $H_2$ adsorption on 4MA@BN:$V_B$. (a) 20$H_2$/4Li@BO:$V_B$, (b) 24$H_2$/4Na@BO:$V_B$, and (c) 24$H_2$/4K@BO:$V_B$.

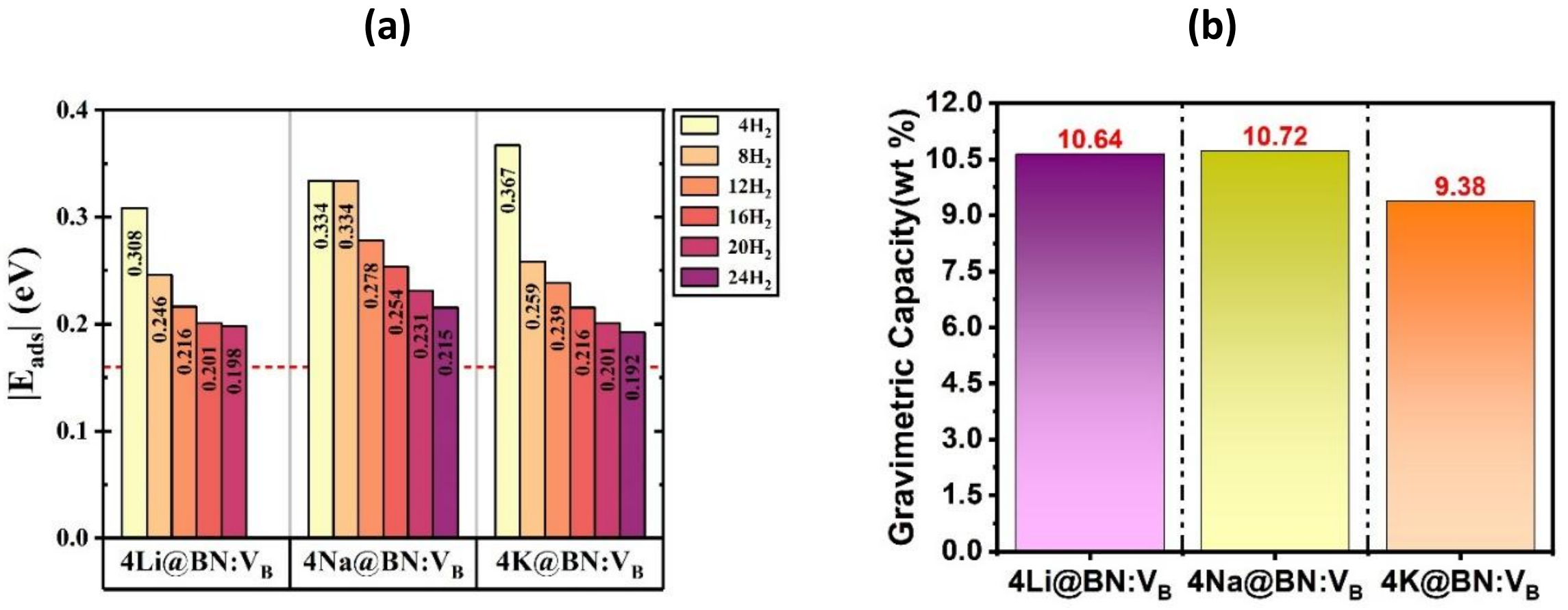

Figure 10: (a) Average adsorption energy per $H_2$ molecule adsorbed on $4MA@BN:V_B$ and (b) the corresponding gravimetric density (wt%).

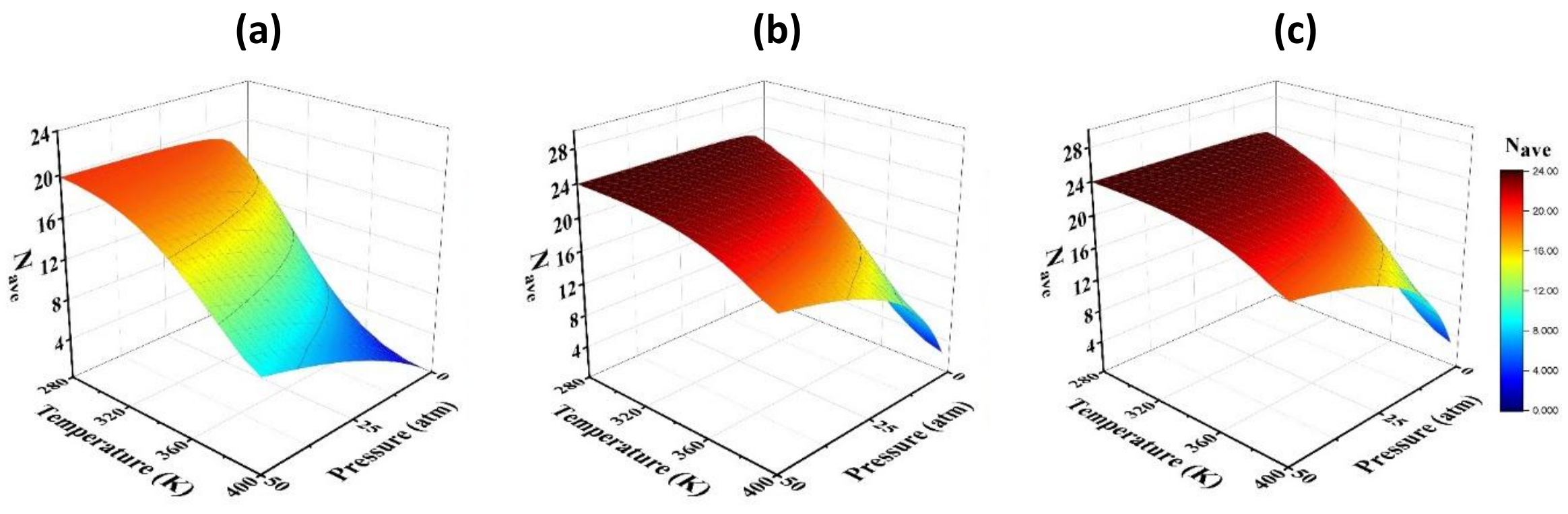

Figure 11: Thermodynamics analysis showing the average number of $H_2$ molecules ($N_{ave}$) adsorbed on (a) 4Li@BN:$V_B$, (b) 4Na@BN:$V_B$, and (c) 4K@BN:$V_B$ as a function of applied temperature and pressure.